\documentclass[10pt, conference, compsocconf]{IEEEtran}

\usepackage{epsfig}
\usepackage{graphics}
\usepackage{amsmath}	   %
\usepackage{shadow}        %
\usepackage{hhline}        %
\usepackage{tabularx}      %
\usepackage{latexsym}      %
\usepackage{times}         %
\usepackage{fancyhdr}      %
\usepackage[hyphens]{url}  %
\usepackage{cite}		   %
\usepackage{verbatim}
\usepackage{xspace}		   %
\usepackage{color}
\usepackage{textcomp}
\usepackage{tcolorbox}

\usepackage{subcaption}
\usepackage{graphicx}

\usepackage{shortcuts}

\renewcommand{\paragraph}{\vspace{3pt}\noindent\textbf}
\begin{document}

\date{}

\setlength{\parskip}{3.5pt}

\thispagestyle{fancy}

\title{How Did That Get In My Phone?\\ \MalApp Distribution on Android Devices}

\author{
  \IEEEauthorblockN{Platon Kotzias\IEEEauthorrefmark{1},
                    Juan Caballero\IEEEauthorrefmark{2},
                    Leyla Bilge\IEEEauthorrefmark{1}\\}
  \IEEEauthorblockA{\IEEEauthorrefmark{1}NortonLifelock Research Group,
                    \IEEEauthorrefmark{2} IMDEA Software Institute}
}

\maketitle

\thispagestyle{fancy}
 
\pagenumbering{arabic}

\begin{abstract}
\noindent 
Android is the most popular operating system with billions of active devices.
Unfortunately, its popularity and openness makes it attractive for \malapps,
\review{i.e., malware and potentially unwanted programs (PUP)}.
In Android, app installations typically happen via the official and 
alternative markets, 
but also via other smaller and less understood alternative distribution 
vectors such as Web downloads, 
pay-per-install (PPI) services, backup restoration, bloatware, and IM tools.
This work performs a thorough investigation on \malapp distribution by 
quantifying and comparing distribution through different vectors.
At the core of our measurements are reputation logs of a large security vendor,
which include 7.9M apps observed in 12M devices between June and September 2019.
As a first step,  
we measure that between 10\% and 24\% of users devices
encounter at least one \malapp, %
and compare the prevalence of malware and PUP.
An analysis of the who-installs-who relationships between \textit{installers}
and child apps reveals that \Play is the main app distribution vector,
responsible for 87\% of all installs and 67\% of \malapp installs, 
but it also has the best defenses against \malapps.
Alternative markets distribute instead 5.7\% of all apps, 
but over 10\% of \malapps. 
Bloatware is also a significant \malapp distribution vector with 
6\% of those installs. 
And, backup restoration is an unintentional distribution vector that may 
even allow \malapps to survive users' phone replacement.
We estimate \malapp distribution via PPI to be smaller than on Windows. 
Finally, we observe that Web downloads are rare, 
but provide a riskier proposition even compared to alternative markets.

\end{abstract}
\section{Introduction}
\label{sec:intro}

Android has become the most popular operating system with over 2.5 billion 
active devices~\cite{numAndroidDevices} and 75\% of the 
mobile device market share~\cite{androidMarketShare}.
A major reason behind Android's success is its open environment that 
allows affordable access to new app developers, 
app distribution through the official \Pl and alternative sources, and  
OS customization by vendors and mobile network operators.
Unfortunately, popularity and openness comes with a cost.
Abusive developers also have easy access to the ecosystem allowing them to 
distribute their \emph{\malapps} to a large number of users~\cite{judy}. 
\review{We use \malapps to 
jointly refer to 
malware (e.g., ransomware, banking trojans) and
potentially unwanted programs (PUP) (e.g., adware, rogueware).}
AV vendors keep reporting ever-increasing numbers of 
\malapp detections and collected \malapp 
samples~\cite{trendmicro2018,mcafee2019,crowdstrike}.

By default, Android only installs apps from the official \Pl, 
but the user can optionally enable installations from other (unknown) sources.
Among those,
alternative markets are popular especially in countries like China where Google
services such as \play are restricted~\cite{ng2014android,wang2018beyond}.
Since app markets are popular and open to any user, prior work has focused on
analyzing them 
\cite{zhou2012hey,ng2014android,androzoo,taylor2017update,wang2018beyond}.
\review{However, the security community lacks a global understanding
about how Android \malapps are distributed. 
While apps are largely distributed through markets, 
other smaller alternative
distribution vectors such as Web downloads, pay-per-install (PPI) services, 
bloatware, backup restoration, and even instant messaging (IM)
should not be ignored. 
We close this gap in the existing research by
investigating how \malapps get distributed into user
devices through different distribution vectors. 
We expect our findings to drive future defenses that protect users when 
installing apps from lesser-known distribution vectors
such as backup restoration and bloatware,
which are still responsible for a significant fraction of \malapp installs. 
We also expect our findings to motivate further research on protecting \Play, 
which despite its defenses, remains by far 
the largest \malapp distribution vector.}

At the core of our measurements are reputation logs that, over the four-month
period between June and September 2019, capture the presence in 12M Android
devices of 7.9M apps (34M APKs)
and the who-installs-who relationship between apps.  Such internal view of what
is installed on user devices, and how it arrived, allows us to answer open
questions such as what fraction of apps are installed through different
distribution vectors, which distribution vectors install more \malapps, which
apps in the same distribution vector (e.g., different markets or browsers)
are riskier compared to each other, and what is the prevalence of \malapp
encounters by users.

\review{Prior works have also leveraged a view of apps 
installed in real user devices.
A recent study analyzed the presence of pre-installed apps in
2.7K Android devices~\cite{preinstalled}. 
In contrast, we examine the distribution of \malapps by pre-installed 
bloatware, and compare it with other distribution vectors.
Furthermore, our device dataset is three orders of magnitude larger
and contains longitudinal logs over a four-month period. 
Another recent work detects stalking apps 
by analyzing 50M Android devices during 2017--2019~\cite{kevincreepware}. 
In comparison, our analysis is performed
on a smaller set of devices in a four-month-period.  However, our analysis
does not focus on one type of threat, but rather covers a large variety of
\malapps distributed through various vectors.
Shen et al.~\cite{shen2016insights} compared rooted and unrooted devices,
measuring the prevalence of five malware classes on 6M Android devices
during April 2015. We also measure prevalence, 
finding, albeit on a four
month period and including all malware and PUP classes, 
significantly higher prevalence.}

As a prerequisite to analyze \malapp distribution, 
we first identify \malapps in the dataset. 
We use the common practice to collect AV detection labels using 
the VirusTotal (VT) online service~\cite{vt} and consider unwanted
any APK flagged by at least a threshold number of 
AV engines~\cite{zhu2020measuring}.
Using those \malapks we measure the number of devices that 
encountered at least one \malapp over the four month analysis period. 
We measure an \malapp prevalence of 10\%--24\%, 
depending on the selected VT threshold.
This range is very conservative as it considers benign all APKs not 
queried to VT or not found in VT. 
Compared to previous studies on Windows malware prevalence on consumer and 
enterprise hosts~\cite{yen2014epidemiological,enterprise,ppipup}, 
this shows that, despite many security improvements provided by the 
Android ecosystem,
the security posture of Android devices with respect to \malapps is not better
than that of Windows hosts.
We also compare the prevalence of malware and PUP finding that the 
prevalence of both categories is almost identical, 
although we identify significantly more PUP samples on the devices.

Then, we examine the who-installs-who relationships between 
\textit{installers} and the child apps they install. 
To compare distribution vectors we classify the installer apps into 12
categories: 
the official \Pl,
alternative markets,
browsers,
commercial PPI,
backup and restore,
IM,  
theme stores, 
file managers, 
file sharing apps, 
bloatware, 
mobile device management (MDM), and
package installers. 
To compare distribution vectors we compute their 
\textit{vector detection ratio} (VDR), 
i.e., the ratio of \malapps installed through that vector over 
all apps installed through that vector.
Below we list our most significant findings on \malapp distribution:

\begin{itemize}

\item \PLAY is the main app distribution vector responsible for 87\% of 
all installs and 67\% of \malinstalls. 
However, its VDR is only 0.6\%, 
better than all other large distribution vectors. 
Thus, \play defenses against \malapps work, 
but still significant amounts of \malapps are able to bypass them, 
making it the main distribution vector for \malapps.

\item Among the remaining installs,
alternative markets are the largest, being  
responsible for 5.7\% of all installs and 10.4\% of \malinstalls. 
However, on average they are five times riskier (3.2\% VDR) than \Play (0.6\%).
Download risk highly varies among alternative markets. 
Some like Amazon's and Vivo's are almost as safe as \play, 
but users of other top alternative markets have up to 19 times higher 
probability of encountering an \malapp.

\item Backup restoration is an unintended \malapp distribution vector 
responsible for 4.8\% of \malinstalls.
Cloning of apps during phone replacement can facilitate \malapps 
to survive phone changes by the user.

\item Bloatware is another surprisingly high distribution vector, 
responsible for 6\% of unwanted installs. 
This is likely due to ad-based monetization by 
device vendors and carriers of the devices they sell. 
Bloatware installers are often privileged, making their removal by 
security tools and users challenging.

\item App downloads from the Web are rare (\textless0.1\% installs),
but have significantly higher risk (3.8\% VDR) than 
downloads from markets, even alternative ones (3.2\%).

\item We provide a very conservative lower bound on commercial PPI service
distribution of 0.2\% of all installs and 0.1\% of \malinstalls and observe 
that such services seem to have improved their filtering of abusive 
advertisers compared to their Windows counterparts.
We also estimate that all PPI activity may be responsible for up to 
4\% of the unwanted app installs.
That upper bound is still significantly lower than the estimate of 
Windows commercial PPI services being responsible 
for over a quarter of PUP installs~\cite{ppipup}. 

\end{itemize}
\review{
\section{Background}
\label{sec:background}

Android apps are distributed as Android application packages (APKs), which are
compressed files that contain the app's code (e.g., DEX files and ELF
libraries), a manifest file, certificates, resources, and other assets.  The
manifest file contains a package name commonly used to identify the
app. The developer is free to choose the package name and
collisions are possible between apps from different developers.  However,
some markets, including the official \Pl, use the package name as a unique
app identifier and therefore do not accept two 
apps with the same package name. For this reason, benign developers 
avoid reusing existing package names.
On the other hand, \malapps may impersonate benign apps by 
selecting the same package name as an app in \play, 
and distributing the impersonating app through alternative markets
(e.g.,~\cite{zhou2012hey,crussell2012attack}).

\subsection{App Signing}
\label{sec:appsigning}

APKs are digitally signed using a private key and they include the signature
and a certificate chain for the corresponding public key, which the Android
framework uses during installation to validate that the APK has not been
modified. In practice, the vast majority of APKs contain only a single
self-signed certificate.
An installed app can only be updated
by another APK with the same package name and only if the new version is signed
with the same private key and has the same certificate as the old version.
To prevent \maldevs from surreptitiously updating the benign apps with
their own versions, it is strictly necessary that developers keep their
private keys confidential.
We use the term \textit{signer} to refer to the entity that owns the private
key that signs an APK.  We identify the APK's signer by either the SHA1 hash of
the certificate %
or by the SHA256 hash of the certificate's public key. %
Both identifiers are essentially equivalent since in an update the Android
framework checks that the hash of the certificates is the same, providing
no incentive to reuse public keys across certificates~\cite{barrera2014baton}. 

Fake apps may impersonate benign apps by using the same package name 
as the benign app
and distributing the fake app through alternative distribution vectors.
Impersonation allows the fake app to inherit the positive reputation of
the benign app. 
Fake apps often correspond to \textit{repackaged}
versions of the benign app with some possibly malicious modifications
(e.g.,~\cite{zhou2012hey,zhou2012detecting,crussell2012attack,chen2015finding}).
Unless the attacker compromises the private
key of the app being impersonated, the fake app will be 
signed with a different private key, and have a different certificate. 
To identify such impersonations and distinguish the
original app from the impersonating app, we track apps in our data using
both their package name and signer.  

\paragraph{Platform keys.}
Building an Android OS distribution requires the publisher to provide four
pairs of public/private keys: \textit{platform}, \textit{test},
\textit{shared}, \textit{media}. 
Among these, the platform key is used to sign
the core Android platform packages. 
APKs signed with a platform key can use
\textit{System} and \textit{SignatureOrSystem}
permissions~\cite{felt2011effectiveness}.
Each device vendor will have at least one platform
key to build its Android images.  Some vendors may use different
platform keys for different devices.  
The Android Open Source Project (AOSP) repository contains default 
platform, test, share, and media 
key pairs, and their corresponding certificates. 
The default AOSP private keys are used to sign Android OS debugging builds,
and should be avoided when building production releases
since the private keys are not really private. 
Signing an APK with an AOSP key
is a well known security issue since any other app also using the AOSP
certificate can update those apps.

\subsection{App Installation}
\label{sec:installation}

This work analyzes the who-installs-who relationships among apps.
In particular, we examine installations where a \textit{parent} app
installs a \textit{child} app. When parent and child are the same app, 
we call it an app update, otherwise we call the
parent app an \textit{installer} and the event an \textit{install}. 
When installing a new app, the Android framework stores the package name of the 
installer, which can then be accessed using method 
\emph{PackageInstaller.getInstallerPackageName}\footnote{\review{Added in API 
level 5 (October 2009). 
Replaced in API level 30 (February 2020) by three methods 
from the \emph{InstallSourceInfo} class: 
\emph{getInitiatingPackageName}, 
\emph{getInstallingPackageName},
\emph{getOriginatingPackageName}~\cite{androidPackageInstaller}.}}.
The installer package is only updated if the app is re-installed from
a different source. 
The installer package may be null if the installer is unknown, 
e.g., when the APK was pre-installed.

Since Android API level 26 (August 2017), a user-level app that wants to 
install another app should declare the 
\url{REQUEST_INSTALL_PACKAGES} permission.
In addition, if the installer is not a trusted source 
(i.e., not a first party market such as \play or the 
device manufacturer market), 
it needs to hold the \emph{install from unknown sources} 
permission\footnote{\review{Prior to API level 26 (August 2017), 
install from unknown sources was a system-wide configuration.}},
which has to be explicitly granted by the user to the 
installer~\cite{androidRequestInstall}.
Even if the installer is authorized to install from unknown sources, 
the user is prompted to authorize the install. 
To perform installs that do not require user consent, 
i.e., \emph{silent installs}, 
the installer must hold the system-level \url{INSTALL_PACKAGES} permission, 
which is only granted to apps signed by the platform key and 
privileged apps explicitly granted that 
permission~\cite{androidPriviledgedPermission}. 

\paragraph{Uninstallation.}
Removing a user-level app
prompts the user to accept the uninstallation. 
Silent uninstallations require the uninstaller to hold the 
system-level \url{DELETE_PACKAGES} permission and 
run in Device or Profile Owner modes~\cite{androidProvisioning}.
Given these requirements, AV engines
prompt the user to uninstall detected \malapps.
Furthermore, system apps 
(i.e., installed under the read-only \url{/system/} directory) 
cannot be uninstalled, only disabled, unless the device is rooted.

\subsection{\MalApps, Malware, PUP}
\label{sec:unwanted}

Malware %
is any software that intentionally 
causes harm to computer systems, networks, and their users.
Some examples of malware classes are %
ransomware, banking trojans, and backdoors. 
In contrast, \emph{potentially unwanted programs} (PUP)~\cite{malwarebytesPUP}, 
also known as \emph{grayware}~\cite{nortonGrayware,andow2016study},
\emph{potentially unwanted applications} (PUA)~\cite{aviraPUP,microsoftPUP},
or \emph{unwanted software}~\cite{googlePUP}, 
are software that, while not outright malicious 
(i.e., not malware) 
still may negatively impact
computer systems, networks, and their users,
e.g., in terms of privacy, 
performance, or 
user experience. %
PUP includes software
that performs abusive advertising (\emph{adware}), 
that does not implement the claimed functionality (e.g., \emph{rogueware}), and 
tools that some users may want to install, but can 
also be abused (e.g., rooting tools).
But, the boundary between malware and PUP is blurry and 
often differs between security vendors~\cite{malwarebytesPUP,nortonGrayware,aviraPUP,microsoftPUP,googlePUP}.
Some classes like \emph{spyware}
(i.e., software the leaks user data) 
are sometimes considered malware and others PUP.

Regardless of such differences, 
which are beyond the scope of this paper and deserve future work,
AV engines alert users about the presence of both malware and PUP 
in their protected devices, 
although PUP treatment may be more lightweight.
For example, mobile AV engines may display more stern and frequent 
notifications to ask the user to uninstall malware 
compared to PUP notifications,
and may even allow the user to disable PUP notifications.

Currently, the security community lacks a term to jointly refer to 
malware and PUP. 
Calling both categories malware raises complaints that PUP is not 
necessarily malicious. 
On the other hand, malware is clearly unwanted. 
In this work, we use \emph{\malapps} to jointly refer to 
Android malware and PUP, 
and separate both categories when needed, e.g., in our prevalence estimations.

}
\begin{table}[t]
\small
\centering
\begin{tabular}{llrr}
\hline
\textbf{Dataset} & \textbf{Data} & \textbf{Full} & \textbf{Subset} \\
\hline
Reputation Logs    & Devices      &  12.2~M \\%
                   & Countries    &  243 \\
\cline{2-3}
                   & APKs                    &  34.6~M & \\ %
                   & Packages                &   7.9~M & \\ %
                   & Signers                 &   4.1~M & \\ %
\cline{2-4}
                    & Unique events       &  2.3~B & \\ %
	                  & Install events      &  1.7~B & 412.6~M \\  %
                    & Installer packages  &  5.4~K &   4.2~K \\ %
                    & Child packages      &  2.8~M &   1.6~M \\ %
\hline
VirusTotal                    & Reports     &  4.6~M & \\ %
                            
\hline
\end{tabular}
\caption{Summary of datasets used.}
\label{tab:datasets}
\end{table}

\section{Datasets}
\label{sec:datasets}

This section details the datasets that lie at the core of our study. 
We use two main datasets summarized in Table~\ref{tab:datasets}.
\emph{Reputation logs} from the \vendor contain information about 
apps installed on 12M Android devices, 
as well as parent-child install relationships among them. 
We query \emph{VirusTotal} (VT) to obtain AV labels for \malapp classification 
and APK metadata such as permissions and certificate info.

\paragraph{Reputation logs.} 
These logs capture metadata about the presence of apps in 12M Android devices.
The dataset does not include the actual apps, but only their metadata. 
These logs are collected from real devices
in use by customers of the \vendor.  The customers opted-in to sharing their
data and the devices are anonymized to preserve the privacy of the customers.
The dataset covers four months that span from June 1st, 2019 to
September 30th, 2019.

\begin{table}
\parbox{.42\linewidth}{
\small
\centering
\begin{tabular}{|r|l|r|}
\hline
 \textbf{\#}
 &\textbf{Country}
 &\textbf{Devices} \\
\hline
1 & United States & 20.7\% \\
2 & India & 17.7\% \\
3 & Japan & 15.3\% \\
4 & Germany & 7.1\% \\
5 & United Kingdom & 5.3\% \\
6 & Brazil & 3.3\% \\
7 & Canada & 2.6\% \\
8 & Australia & 2.5\% \\
9 & France & 2.3\% \\
10 & Netherlands & 2.1\% \\
11 & Italy  & 1.9\% \\
12 & Spain & 1.6\% \\
13 & Poland & 1.3\% \\
14 & Belgium & 1.0\% \\
15 & Russia & 0.9\% \\
\hline
\end{tabular}
\caption{Top 15 countries by devices.}
\label{tab:geo}
}
\hfill
\parbox{.42\linewidth}{
\small
\centering
\begin{tabular}{|l|r|}
\hline
 \textbf{Vendor}
 &\textbf{Devices} \\
\hline
Samsung & 40.5\% \\
Xiaomi & 8.6\% \\
Motorola & 7.0\% \\
LYF & 4.9\% \\
Huawei & 4.6\% \\
Sony & 4.6\% \\
LGE & 4.3\% \\
Lenovo & 4.3\% \\
Sharp & 2.6\% \\
Asus & 1.9\% \\
Fujitsu & 1.8\% \\
HMD Global & 1.3\% \\
OnePlus & 1.1\% \\
Oppo & 1.1\% \\
Google & 1.0\% \\
\hline
\end{tabular}
\caption{Top 15 device vendors by devices.}
\label{tab:manufacturer}
}
\end{table}

The dataset contains devices in 243 country codes~\cite{cc_iso}.
thus covering nearly all countries in the world, save a few exceptions 
like North Korea.
The top 15 countries by number of devices are shown in Table~\ref{tab:geo}. 
These 15 countries cover 89\% of the devices, but the distribution is 
long-tailed.
The dataset is skewed towards North America, Europe, and Japan where 
the \vendor has a larger market share.  
Of the 20 largest countries by population we see that 
China, Indonesia, Pakistan, Nigeria, and Bangladesh are underrepresented, 
but we still have tens of thousands of devices in China and several thousands 
in the rest.

The dataset includes devices from over 3K device vendors.
Table~\ref{tab:manufacturer} shows the top 15 vendors by devices 
in the dataset. 
Samsung is the dominant device vendor with over 40\% of the 
devices, followed by Xiaomi (8.6\%) and Motorola (7.0\%).
Again, the distribution is long-tailed with only 14 vendors having more 
than 1\% of the devices.

Each device in the dataset regularly queries a cloud-based reputation system
to obtain the reputation %
for the APKs installed in the device.
The query includes file metadata such as 
APK hash, APK package name, the signer key 
(i.e., the SHA256 of the public key in the APK's certificate), and 
optionally the name of the parent package that installed the APK.
The response includes a reputation score, 
which is one of the inputs, but not the only one, 
used by the {\vendor}'s AV engine to make a determination about an APK.
Since the reputation score is proprietary, 
we avoid using it to make our approach replicable.
We only use it to select samples with low reputation to query to VirusTotal. 

The \client may query the same APK at different times. 
To remove duplicated events, 
for each unique tuple of 
an anonymized device identifier,
APK's SHA256 hash,
APK's package name,
APK's signer key, and
APK's parent package name (potentially null), 
we obtain the earliest date when the tuple was queried to the reputation system.
The dataset comprises of 2.3B such unique events with 
34.6M APKs from 7.9M packages using 4.1M certificate chains.

The \client queries Android's Package Installer\footnote{\review{Using the \emph{PackageInstaller.getInstallerPackageName} method.}}
to obtain the name of the parent package for each installed APK. 
If the parent package is known, it
is included in the query to the reputation server. However, in some cases
parent packages might be unknown to the Package Installer. Some examples are
apps that come preinstalled on the device
and sideloaded apps installed via the Android Debug
Bridge (ADB) and for which the user did not provide an installer
package name.
Of the 2.3B unique events in the dataset, 
75\% correspond to installations 
(i.e., have a parent package different from the child package), 
24\% correspond to updates
(i.e., same parent and child package), and
1\% have no parent package information.
The 1.7B install events contain 
5.4K parent (installer) packages and 2.8M child packages.

\review{
The interplay between the AV engine and the reputation log collection 
is as follows.
The APK reputation is part of the decision made by the \client.
APKs are queried to the reputation server prior to making a determination.
Thus, apps flagged by the \client will appear in the 
reputation logs for the device.
Upon detection, if the app is classified as malware, the \client 
displays a large warning and asks for permission to uninstall it. 
If classified as PUP, a notification explains the risks 
to the user and how to uninstall it.}

\paragraph{VirusTotal.}
We query the hash of APKs in VirusTotal (VT)~\cite{vt},
an online service that analyzes files and URLs submitted by users
using a large number of security tools.
VT offers a commercial API that given a file hash returns 
file metadata and the list of detection labels assigned by a 
large number of AV engines used to scan the file.
Unfortunately, given VT's API restrictions, we could not query all 34.6M APKs. 
We queried 
all parent APKs that performed at least one installation
(i.e., all installers), 
the 10 most prevalent APKs for each signer,
all the APKs with negative reputation, and 
a subset of the APKs with positive reputation.
This resulted in VT reports for 4.6M APKs. %
We use the AV labels in the VT reports as an input to our \malapp 
identification and classification.
Since we only have VT reports for 13\% of all APKs, 
our \malapp prevalence results is a lower bound.
We also use the VT reports to obtain APK metadata such as 
certificate information, used to analyze APK ownership, 
and permissions declared in the manifest, 
used to identify installers that can perform silent installations 
without user consent.
Having APK metadata for 13\% of samples does not affect our results 
because we only use the certificate information for analysis of 
selected samples and permissions for installers.
In both cases, we have queried the necessary APKs. 

\paragraph{\Pl.} We check if an app found in user devices is available in
Android's official market by trying to download its public webpage using the
app's package name.  For apps in \play, 
we obtain metadata such as its category. 
We queried all 7.9M package names during February 2020. 
Of those, 24\% (1.9M) were present at that time in \play. %
More may have been available in the past, 
but have since been removed~\cite{wang2018android}. 
The rest may be distributed only through 
alternative distribution vectors, or may come pre-installed.

\section{Approach}
\label{sec:approach}

This section first describes data challenges we had to overcome 
and then explains how we identified platform keys and
categorized installers.

\paragraph{Obtaining parent information.}
The reputation logs contain the 
package name of the parent APK, but not the parent APK's hash 
or public key. 
This is problematic because benign apps could be impersonated by 
\malapps, misleading us into assigning \malinstalls to benign 
installers.
To avoid this, for each install event, 
we scan all reputation queries from that device during the 4 months, 
extracting those that queried an APK whose package name corresponds to the 
parent in the install event (i.e., the parent we look for appears as child). 
If such reputation queries exist, we sort them by decreasing 
time difference from the install event and assign 
the APK's hash and public key from the closest event as the parent .

We apply this procedure on the 1.7B install events in the 
reputation logs (\textit{Full} column in Table~\ref{tab:datasets}),
recovering parent information for 
24\% (412.6M) of the install events. %
This 24\% subset of install events covers 
78\% (4.2~K) of the installer packages and 57\% of the child packages 
in the full install events, 
as summarized in column \textit{Subset} in Table~\ref{tab:datasets}.
Recovery failures are likely due to each APK 
being assigned a time-to-live (TTL) indicating when to re-query its reputation. 
For APKs positively benign, the TTL may be large enough so that the 
install event happens before our study period and the device never re-queries 
(or leaves the dataset before).
\review{Thus, the 24\% install events may be skewed towards \malinstallers, 
which may bias VDR absolute numbers.
We avoid this bias by computing the relative VDR and by confirming 
results do not significantly change when computed on all 1.7B install events.}

We use the Full dataset of 34.6~M APKs installed on 12M devices 
to analyze \malapp encounters in Section~\ref{sec:threats}. 
We use the Subset of 24\% install events to analyze distribution vectors 
in Section~\ref{sec:installers}.

\paragraph{Identifying platform keys.} To identify platform keys in the
reputation logs we first obtain from the AOSP repository a list of 65 package
names that are part of the Android OS.  Then, we search for keys in the
reputation logs that satisfy the following properties: signs
\url{com.android.phone} (a core Android package) and 
signs at least ten AOSP packages.
Third, for each of those candidate keys we identify the top 10 packages signed
by the key.  If at least half of those packages are present in the list of 65
AOSP packages we keep the key, otherwise we remove it. 
Finally, we examine the certificate information.
If the subject DN mentions a specific vendor, 
and we are able to find a webpage for the vendor,
we keep the key. Otherwise we remove it. 
For 10\% of the examined keys we could not identify a vendor due to 
a generic Subject DN.
This verification is manual, so we restrict it to keys that appear in 
more than 1K devices. 
However, thanks to this verification, 
we are confident that the resulting keys are platform keys. 
Using this procedure, we identified 201 platform keys belonging to 80 device
vendors or OS publishers.  The highest number of platform keys is 57 for
Motorola that uses separate keys for different devices.
Those 201 platform keys appear in over 6M (50\%) devices. 
Thus, when we say a key is a platform key we are confident about it, 
but we may miss that some keys (e.g., for less prevalent vendors) 
are indeed platform keys.

\paragraph{Installer categorization.}
To analyze how apps are distributed to the devices, 
we manually classify the installers into 12 categories 
that correspond to distribution vectors: 
the official \Pl, 
alternative markets, 
browsers, 
commercial PPI, 
backup and restore, 
IM, 
theme stores, 
file managers, 
file sharing apps.
bloatware, 
mobile device management (MDM), and
package installers. 
Bloatware~\cite{mcdaniel2012bloatware} corresponds to apps signed by a 
device vendor or a carrier with unclear functionality,
i.e., they do not belong to any of the other categories. 
Bloatware typically comes pre-installed, although it could be  
installed later as well.
MDM apps enable the administration of corporate mobile devices, 
e.g., ensuring corporate apps are installed and 
the corporate security policy is configured.
\review{Package installers are apps that enable installing APKs. 
They include implementations of Android's Package Installer module
(e.g., \url{com.android.packageinstaller},
\url{com.google.android.packageinstaller},
\url{com.samsung.android.packageinstaller})
as well third-party APK installers
(e.g., \url{com.apkinstaller.ApkInstaller}, \url{com.aefyr.sai}).
}
We also add an \textit{Other} category that comprises of apps 
that we can classify but do not correspond to any of the 12 
expected distribution vectors such as games, video players, and news.
For installers available in \play we leverage their app description 
since market categories are too coarse-grained. 
However, only 12\% of the installers were in \play when we queried them 
in February 2020. 
For the rest, we need to examine sources such as alternative markets, 
results from Web searches, and forums.
This process is quite challenging 
for the long tail of less popular installers.
Overall, out of the 4.2K installers, only 665 install at least one \malapp. 
We focus on those and are able to classify 622 (95\%). 
Those 662 include the most prevalent installers so we classify
96.3\% of the 412.6M install events.
The largest category is alternative markets with over one hundred installers. 
The results of the categorization are detailed in Section~\ref{sec:vectors}.

\section{\MalApp Encounters}
\label{sec:threats}

This section reports on \malapp 
encounters that affected the 12M devices in 
the full reputation logs.
Section~\ref{sec:prevalence} measures the prevalence of 
\malapps on the user devices. 
Section~\ref{sec:classification} describes the most common 
families in our dataset, and  
Section~\ref{sec:malsigners} details the top 
signers behind the \malapps. 

\begin{figure}
	\includegraphics[width=.9\columnwidth]{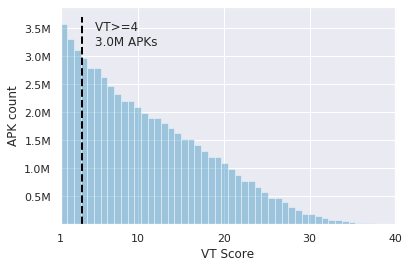}
	\caption{\Malapks per VT detection threshold.}
  \label{fig:vtt_apk}
\end{figure}

\begin{figure}
	\includegraphics[width=.9\columnwidth]{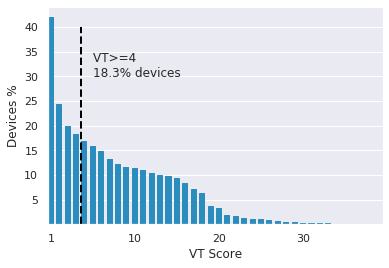}
	\caption{\Malapp prevalence per VT detection threshold.}
  \label{fig:vtt_prevalence}
\end{figure}

\subsection{\MalApp Prevalence}
\label{sec:prevalence}

We measure the prevalence of \malapps on user devices, 
i.e., the fraction of user devices that had an \malapp encounter 
throughout the four months analysis period.
For this, we first identify \malapps installed on user devices. 
Then, we measure their prevalence across the 12M user devices. 
A common practice to identify \malapps
is to collect their AV detection labels using VT and consider 
any file flagged (i.e., assigned a non-empty label) by
at least a threshold number of AV engines~\cite{zhu2020measuring}. 
A higher threshold reduces false positives due to a few AV engines 
making an incorrect determination, but may increase false negatives.
As explained in Section~\ref{sec:datasets}, 
we could not query all 34.6M APKs due to VT API restrictions, 
but were able to collect VT reports for 13\% (4.6M) of all APKs. 
Figure~\ref{fig:vtt_apk} shows the number of \malapks in our dataset 
depending on the selected VT $\ge$ $t$ threshold.
The number of \malapks decreases as the threshold increases. 

We use the set of \malapks obtained at each threshold value 
to compute the device prevalence, 
i.e., the fraction of devices where those \malapks were installed.
Threshold selection varies among different works~\cite{zhu2020measuring}. 
Thus, we provide the prevalence at all threshold values in 
Figure~\ref{fig:vtt_prevalence}.
The prevalence quickly decreases from t=1 to t=3, 
and then at around one percentage point per step increase until t=19.
Recent work has shown that threshold values between two and 14 are good 
for stability and for balancing precision and recall~\cite{zhu2020measuring}.
Thus, the \malapp prevalence in our dataset ranges between 
24.3\% (t=2) and 10.0\% (t=14). 

\begin{table}[t]
\centering
\begin{tabular}{|l||r|r|r|}
\hline
\textbf{Value}   & \textbf{\MalApps} & \textbf{PUP} & \textbf{Malware}  \\
\hline
Devices &  2.2~M (18.3\%) & 1.3~M (11.1\%)  & 1.4~M (11.2\%) \\  
APKs    &  3.0~M  (8.6\%) & 1.8~M  (5.1\%)  & 1.2~M  (3.4\%)  \\
\hline
 \end{tabular}
\caption{\Malapp prevalence at selected t=4.}
\label{tab:prevalence}
\vspace{-12pt}
\end{table}

For the rest of the paper we need to set a threshold value, 
so that the set of \malapks is fixed. 
We select t=4 as our threshold.
This value falls in the range recommended in~\cite{zhu2020measuring} and 
has been used in closely related works (e.g.,~\cite{malsign,ppipup}).
Using this threshold, there are 3.0M \malapks and the 
prevalence is 18.3\%, as summarized in Table~\ref{tab:prevalence}.
Clearly, this estimate is very conservative as it considers benign all APKs 
that were not queried to VT, or were not found in VT, 
or were flagged by less than four AV products.
We believe this is a lower bound for prevalence.
Among the devices with at least one \malapp encounter at t=4, 
the median is 2.0 \malapps per device
(avg=5.0, std=1497.0).
Figure~\ref{fig:malapps_per_device} in the Appendix 
details the distribution.

Table~\ref{tab:prevalence} also provides the split between malware and PUP APKs,
according to the AVClass malware labeling tool~\cite{avclass}  
(see Section~\ref{sec:classification}).
It shows that 60\% of the \malapks at t=4 are considered PUP and 40\% malware.
However, malware prevalence (11.2\% of all 12M devices) is almost the same as 
PUP prevalence (11.1\%), 
indicating the presence of some high prevalence malware. 
The devices typically encounter only malware or PUP, 
but 490K devices encounter both types.

\paragraph{Comparison with prior work.}
Shen et al~\cite{shen2016insights} measured the fraction of devices 
with an AV detection over 6M Android devices in April 2015. 
Those detections belonged to five categories: 
trojan, infostealer, backdoor, hacktool, spyware. 
Unfortunately, they only provide the fraction of rooted and non-rooted devices
with at least one detection of each category.
Each number is in the range 0.10\%--0.71\%, but we cannot infer the 
total prevalence, making the comparison difficult.
Still, our 10\%--24\% prevalence range seems significantly higher. 
This is likely due to our measurement including also PUP and 
covering four months.
In addition, the last four years may have seen an 
increase in \malapps as Android became the dominant OS 
in user devices.
We also examined recent threat reports from
\vendors~\cite{trendmicro2018,mcafee2019}. However, those reports do not
measure prevalence on devices, but rather number of detections and number of
new samples discovered, which cannot be accurately compared.
Other works have measured malware and PUP prevalence on Windows enterprise
hosts~\cite{yen2014epidemiological,enterprise}. Some works~\cite{enterprise} operated on a very different time period (three years
versus four months in our study), making it difficult to compare prevalence
estimates.
However, Yen et al.~\cite{yen2014epidemiological} observed that 15\% of hosts
in a large enterprise encountered at least one detection by a 
specific AV engine over a four-month period in 2014.
Yen et al. call the flagged samples malware, but the authors confirmed us they 
included all detections by the \client.
That prevalence falls in our 10\%-24\% range, 
also computed on a four-month period although in mid-2019, 
and is slightly lower than the 18.3\% prevalence with the selected threshold. 

\review{
\paragraph{Takeaway.} Even the very conservative estimate of 18.3\%
indicates that Android devices have an \malapp prevalence 
rate similar %
and possibly slightly higher than Windows (enterprise) hosts.
Thus, despite many security improvements provided by the Android ecosystem
(e.g., OS app isolation, OS permission model, official market) the
security posture of Android devices with respect to \malapps 
does not seem better
than that of Windows (enterprise) hosts.
}

In the remainder of the paper, \malapps
refers to the 3M APKs at t=4,  
i.e., flagged by at least 4 AV engines. 

\begin{table*}[t]
\caption{Top 10 \malapp signers by prevalence.}
\footnotesize
\centering
\begin{tabular}{|r|l|l|r|r|r|r|r|r|}
\hline
 \textbf{\#}
 &\textbf{Certificate Thumbprint}
 &\textbf{Name}
 &\textbf{Devices}
 &\textbf{Packages}
 &\textbf{APKs}
 &\textbf{Unw. APKs}
 &\textbf{SDR}
 \\
\hline
1  & 61ed377e85d386a8dfee6b864bd85b0bfaa5af81 & AOSP Test                &  1,811,862 & 480,639 & 1,444,423 & 255,393 & 17.7\% \\ %
2  & 9edf7fe12ed2a2472fb07df1e398d1039b9d2f5d & O=Qbiki Networks         &  328,286   & 75,537  & 134,488   & 27,073  & 20.1\% \\ %
3  & 27196e386b875e76adf700e7ea84e4c6eee33dfa & AOSP Platform            &  197,602   & 22,782  & 355,841   & 16,489  & 4.6\%  \\ %
4  & 3246bde9e58a7e0cdf779a7b403581ba958361c3  & O=Outfit7 Ltd.           &  156,932   & 40      & 1,363     & 219     & 16.1\% \\ %
5  & 614d271d9102e30169822487fde5de00a352b01d & OU=gsr                   &  68,239    & 21,758  & 26,060    & 7,277   & 27.9\% \\ %
6  & ac640e8372e429f9894a5e1dff1081e223aa94e3  & CN=1mobile               &  66,508    & 20      & 17,541    & 2,045   & 11.7\% \\ %
7  & bc87c82cd2886a4e07e1f2e1156ddc9b2c467dc8  & O=NetDragon              &  60,171    & 13,176  & 31,311    & 10,806  & 34.5\% \\ %
8  & 5d08264b44e0e53fbccc70b4f016474cc6c5ab5c  & CN:Android Debug         &  52,402    & 1,010   & 19,066    & 9,696   & 50.9\% \\ %
9  & 971da0d8842f7539c666f87b74676c4548c26341  & CN=iMobLife              &  46,609    & 230     & 1,849     & 130     & 7.0\%  \\ %
10 & 6d2aa36c370d8b6156dba70798a8c6c728265404  & CN=Pravesh Agrawal       &  28,302    & 6,104   & 10,648    & 6,343   & 59.6\% \\ %

\hline
\end{tabular}
\label{tab:malpub}
\end{table*}

\subsection{\MalApp Signers}
\label{sec:malsigners}

We use the certificate's public key hash to identify the signer, 
i.e., the entity that signs an APK. %
For each of the 4.1M signers in the full reputation logs, 
we compute the \textit{signer detection ratio} (SDR),
i.e., the fraction of \malapks 
it signs over the total number of APKs it signs.
The SDR allows to identify signers with a significant fraction of \malapks. 
The distribution is bimodal, 
86\% of the signers have 0\% SDR 
and 13\% have 100\% SDR, with only 1\% in between.
We consider a signer unwanted if it has SDR $\ge$ 4\% and signs 
at least 100 \malapks. 
We experimentally chose these thresholds to minimize flagging signers
with a small number of APKs compromised in a short time period 
(e.g., using an abusive SDK in some versions), 
who later on addressed the issue.
Using these thresholds, we flag 146 (0.003\%) of all signers as unwanted, 
with only 70 of those being present in over 1K devices.

Table~\ref{tab:malpub} shows the top 10 unwanted signers 
by device prevalence.
For each signer, the table shows 
ranking; 
certificate thumbprint; 
a name that corresponds to a well known key or a distinguishing 
substring of the certificate's Subject DN; 
number of devices where APKs from the signer are installed, 
number of distinct package names of APKs signed by the signer, 
number of APKs signed by the signer, 
number of \malapks among those, and 
SDR.

Six of the 10 signers correspond to generic certificates that 
sign apps from multiple developers.
Ranks 1 and 3 correspond to default AOSP keys. 
Signing an APK with an AOSP key is a well known security issue 
since the private keys are known and any other apps signed with the same key
could perform an update. 
There are 1.83M (1.5\%) devices with APKs signed by any of
the AOSP keys. 
Since \play does not allow apps signed with these keys to be uploaded, 
these apps must have been installed from alternative sources.
APKs signed with these keys include popular package names 
such as MineCraft and WhatsApp, 
indicating impersonation of benign apps by \maldevs.
Using the AOSP keys facilitates the \maldevs to hide among the crowd 
of other developers also using them. 
Prior work has observed the use of AOSP keys in 
custom firmware images~\cite{zhou2012dissecting,zheng2014droidray}. 
Our work differs in reporting their abuse and the 
fraction of devices affected by them. 
At rank 8 there is an Android Debug certificate,  
used by Android Studio to sign applications during debugging.
Similar to the AOSP certificates, such certificates should not be 
used outside debugging, and applications using them cannot be uploaded to 
\play.

Generic certificates at ranks 2, 5, and 10 correspond to the 
SeattleCloud~\cite{seattleCloud}, 
AppsGeyser~\cite{appsGeyser}, and 
WordPress2Apk~\cite{wp2Apk} 
online app generators (OAGs), respectively. 
\review{OAGs automate app development, 
lowering the technical skill required by app developers, 
and may offer publishing the produced apps to 
\Play~\cite{oltrogge2018rise}.
Our results indicate OAGs publish \malapps from their clients, 
possibly due to limited vetting of the apps they (are paid to) publish.}
Identifying such generic certificates is fundamental to avoid false positives 
in detection systems that leverage signer reputation, 
as well as when attributing apps to their owners~\cite{attribution}.

The remaining four signers correspond to specific \maldevs. 
Both 1Mobile and Netdragon do not have any apps in the Play market.
1Mobile publishes an alternative market of the same name.
The 1Mobile market app (\url{me.onemobile.android}) has 17,466 APKs, 
an unusually high number of versions possibly indicating polymorphism 
to bypass detection.
NetDragon sold part of its business to Baidu in 
July 2013~\cite{netDragonBaidu}. 
We believe the reason why NetDragon appears in the list is 
the 2015 discovery of a backdoor in Baidu's MoPlus SDK~\cite{moplusSdk}. 
Both Outfit7 and iMobLife have multiple apps in the Play market. 
Surprisingly, iMobLife uses a large number of Play developer accounts 
to distribute their performance optimization and mindfulness apps. 
The accounts include 
\textit{iMobLife Inc.};
\textit{AIO Software Technology CO., Ltd.};
\textit{Daily Yoga Culture Technology Co., Ltd};
\textit{HK SMARTER MOBI TECHNOLOGY CO.,LIMITED};
\textit{SM Health Team}; and 
\textit{The Unexplainable Store\textregistered}.

\review{
\paragraph{Takeaway.} 
In addition to APK polymorphism, \maldevs also leverage developer account
polymorphism, which provides isolation between apps in different
accounts.
In this way, even if some of the apps are removed by the market,
others could remain available.
We observe AOSP keys being abused to sign \malapps and being present in 
1.83M devices. 
Moreover, we also observe online app generators 
being abused to generate and publish \malapps.
}

\begin{table}[t]
\centering
\footnotesize
	\begin{tabular}{|l|ll|r|r|r|r|}
\hline
	 \textbf{Family} & \multicolumn{2}{c|}{\textbf{Type}} & \textbf{Devices} & \textbf{APK} & \textbf{Pkg.} & \textbf{Sig.} \\
  \hline
		necro           & Mal  & Dropper       & 680K   & 985  & 288   & 608 \\ %
		jiagu           & PUP  & Tool          & 577K   & 248K & 82K   & 49K \\ %
		hiddad          & Mal  & Adware        & 225K   & 128K & 11K   &  9K \\ %
		smsreg          & PUP  & SMS           & 179K   & 150K & 79K   & 49K \\ %
		revmob          & PUP  & AdLibrary     & 169K   &  62K & 54K   & 13K \\ %
		inmobi          & PUP  & AdLibrary     & 131K   &  26K & 24K   &  5K \\ %
		leadbolt        & PUP  & AdLibrary     & 127K   &  37K & 28K   & 10K \\ %
		datacollector   & PUP  & Infostealer   & 123K   &   9K &  4K   &  3K \\ %
		autoins         & PUP  & Infostealer   & 112K   &  14K &  6K   &  3K \\ %
		anydown         & PUP  & Adware        & 109K   &  24K &  18K  &  5K \\ %
		mocen           & Mal  & Adware        & 94K    &  753 &  248  & 132 \\ %
		airpush         & PUP  & AdLibrary     & 91K    &  64K &  39K  & 15K \\ %
		dnotua          & PUP  & Riskware      & 80K    &  50K &  32K  &  6K \\ %
		secapk          & PUP  & Tool          & 72K    &  25K &  14K  &  6K \\ %
		domob           & PUP  & AdLibrary     & 70K    &  18K &  10K  &  3K \\ %
		appsgeyser      & PUP  & AppGen        & 59K    &  14K &  13K  & 161 \\ %
		dowgin          & PUP  & Adware        & 58K    &  32K &  25K  &  9K \\ %
		dianjin         & PUP  & AdLibrary     & 58K    &   8K &   5K  & 913 \\ %
		utilcode        & PUP  & Tool          & 50K    &   3K &  785  & 678 \\ %
		secneo          & PUP  & Tool          & 50K    &  13K &   5K  & 3K \\ %
	\hline
 \end{tabular}
	\caption{Top 20 families by device prevalence.}
\label{tab:familiesbyhosts}
\end{table}

\subsection{Family Classification}
\label{sec:classification}

To understand the most prevalent threats affecting Android devices, 
we classify the \malapks into families. 
For this, we feed the AV labels from the 4.6M VT reports to 
the AVClass malware labeling tool~\cite{avclass}.
AVClass outputs the most likely family name for each sample and also
classifies it as malware or PUP based on the presence of PUP-related keywords
in the AV labels (e.g., \textit{adware}, \textit{unwanted}).
Overall, AVClass labels 2.4M (76\%) of the APKs 
belonging to 2.9K families.
For the remaining 700K samples with a VT report,
no family was identified as either they were not detected by 
any AV engine or their AV labels were generic.

Table~\ref{tab:familiesbyhosts} shows the 20 most prevalent malware and PUP
families identified by AVClass. 
PUP clearly dominates malware with 17 versus three families.
Most PUP families are related to abusive advertisement including 
six advertisement libraries
(\textit{inmobi}, \textit{leadbolt}, \textit{airpush}, \textit{domob}, 
\textit{dianjin}), 
four adware families
(\textit{hiddad}, \textit{anydown}, \textit{mocen}, \textit{dowgin}), 
and one app generator that monetizes through advertisement 
(\textit{appsgeyser}).
The ad libraries are added by other
applications and their behaviour varies from displaying in-app ads like
\textit{inmobi} and \textit{leadbolt}, to more aggressive techniques used by
\textit{dianjin} and \textit{airpush} such as ads in the system notification
bar or shortcut ads on the home screen or in the application list. 
Some ad libraries also collect personal identifiable information 
(e.g., GPS coordinates) and track users using permanent identifiers 
(e.g., IMEI), 
which violates Google Play policies~\cite{googleids}. 
Among the PUP families there are also three tools used for obfuscating 
mobile apps (\textit{jiagu}, \textit{secapk}, \textit{secneo}).
Obfuscation tools are commonly used by malware, 
but may be used also by benign software causing 
false positives~\cite{aghakhani2020malware}.

The most prevalent malware family is \textit{necro}, a trojan dropper that
infected over 680K devices.  Necro has been observed embedded in popular
applications available in \play such as CamScanner~\cite{necro}.  According to
the CamScanner developers the malware made it into their app through the third
party advertising SDK provided by AdHub~\cite{camScanner}.
The table also includes two information-stealing families
(\textit{datacollector}, \textit{autoins}).
Other notorious malware families outside the top 20 are the \textit{triada}
rootkit (29K devices)~\cite{triadaKaspersky,triadaInjection},
the \textit{wroba} banking trojan (21K)~\cite{wroba}, and 
\textit{agentsmith} (16K) that 
replaces installed apps such as WhatsApp with modified versions that show
fraudulent ads~\cite{agentsmith}.
We also search for ransomware and other banking trojans in our dataset.
We identify 11 ransomware families affecting in total 31K devices including
\textit{svpeng}, \textit{congur}, and \textit{jisut}.  This very modest
prevalence matches industry reports that show ransomware decreasing in the wild
after 2017~\cite{trendmicro2018, esetransom2018}.
We also identify 17 banking trojans affecting in total 30K devices.  The most
prevalent families are \textit{wroba}, \textit{hqwar}, and \textit{asacub}.
Industry reports mention that banking trojans samples increase over time,
especially after 2018~\cite{trendmicro2018,mcafee2019}.  Their small prevalence
in our dataset can be potentially explained by banking trojans being heavily
polymorphic and AVs not always being able to assign them non-generic labels.

\review{
\paragraph{Takeaway.} The higher PUP sample prevalence in user devices 
observed in Section~\ref{sec:prevalence} also manifests in the largest 
families being PUP.
Top PUP families are mostly ad-related and use popular obfuscation tools 
for protection. 
The largest malware families are information stealers, but we also 
observe rootkits and ransomware in tens of thousands of devices.
}

\begin{table*}[t]
\small
\caption{\review{Summary of app distribution.}}
\centering
\begin{tabular}{|l|r|r|r|r|r|r|r|r|r|r|r|}
\cline{2-10}
\multicolumn{1}{c|}{} & \multicolumn{2}{c|}{{\bf Installs}} & \multicolumn{5}{c|}{\bf Installer} & \multicolumn{2}{c|}{\bf Children} \\
\hline
 \textbf{Vector}
 &\textbf{All}
 &\textbf{Unw.}
 &\textbf{All}
 &\textbf{Unw.}
 &\textbf{Plat.}
 &\textbf{Pkg.}
 &\textbf{Sig.}
 &\textbf{Pkg.}
 &\textbf{Sig.}
 &\textbf{VDR}
 &\textbf{RVDR}
\\
\hline
\hline
Playstore      &  87.2\%          & 67.5\%          & 10  & 3  & 0  & 2   &  9  & 1.2M  & 816K   & 0.6\%  & 1.0 \\
Alt-market     &  5.7\%           & 10.4\%          & 102 & 31 & 15 & 87  &  67 & 128K  & 77K    & 3.2\%  & 5.3 \\
Backup         &  2.0\%           &  4.8\%          & 49  & 2  & 24 & 31  &  39 & 528K  & 355K   & 0.9\%  & 1.5 \\
Pkginstaller   &  0.7\%           & 10.5\%          & 79  & 5  & 25 & 11  &  74 & 197K  & 127K   & 2.4\%  & 4.0 \\
Bloatware      &  0.4\%           &  6.0\%          & 54  & 2  & 28 & 37  &  41 & 2.1K  & 1.3K   & 1.2\%  & 2.0 \\
PPI            &  0.2\%           &  0.1\%          & 21  & 0  & 2  & 20  &  11 & 1.5K  & 1.3K   & 0.3\%  & 0.5 \\
Fileshare      & \textless0.1\%   & \textless0.1\%  & 13  & 3  & 4  & 13  &  11 & 8.8K  & 7.4K   & 1.3\%  & 2.1 \\
Themes         & \textless0.1\%   & \textless0.1\%  & 2   & 0  & 2  & 2   &  2  & 634   & 14     & 0.3\%  & 0.5 \\
Browser        & \textless0.1\%   & \textless0.1\%  & 47  & 4  & 3  & 40  &  38 & 4.8K  & 3.3K   & 3.8\%  & 6.3 \\
MDM            & \textless0.1\%   & \textless0.1\%  & 7   & 1  & 1  & 7   &  6  & 766   & 489    & 0.3\%  & 0.5 \\
Filemanager    & \textless0.1\%   & \textless0.1\%  & 58  & 11 & 9  & 32  &  43 & 6.6K  & 4.7K   & 2.6\%  & 4.3 \\
IM             & \textless0.1\%   & \textless0.1\%  & 13  & 2  & 0  & 10  &  11 & 2K    & 1.2K   & 2.9\%  & 4.8 \\
\hline
Other          & \textless0.1\%   &          0.3\%  & 151  & 68   & 28  & 125  &  98   & 9.1K & 5.3K & 3.9\% & 6.5 \\
Unclassified   & 3.7\%            & \textless0.1\%  & 3.5K & 2.4K & 386 & 3.3K &  814  & 91K  & 16K  & \textless0.1\%  & 0.1 \\ %
\hline
All            & 100.0\%          & 100.0\%         & 4.2K & 2.5K & 79  & 3.6K &  1.0K & 1.6M & 992K & 1.6\% & 2.6 \\
\hline
\end{tabular}
\label{tab:distr_vectors}
\end{table*}

\section{\MalApp Distribution}
\label{sec:installers}

In this section, we investigate \malapp installation vectors,
i.e., how \malapps ended up on the devices. 
To this end, we use the Subset dataset in Table~\ref{tab:datasets} 
with 412M install events, corresponding to the 24\% install 
events for which we could recover parent information, as explained in
Section~\ref{sec:approach}.
We uniquely identify an installer by the pair of its package name and its
signer. That way we can differentiate \malinstallers that impersonate
(i.e., use the package name of) a benign installer, as well as apps that
have multiple signers, e.g., system apps that are signed by different 
device vendors.
As explained in Section~\ref{sec:prevalence}, we consider unwanted 
any APK flagged by at least 4 AV engines.
For each installer, we calculate the \textit{installer detection ratio} (IDR),
i.e., the fraction of unwanted APKs it installs over the total number of
APKs it installs.
We also compute the \textit{vector detection ratio} (VDR) 
as the fraction of unwanted APKs installed using a distribution vector
(e.g., alternative markets, browsers)
over all APKs installed through that vector.

\subsection{Distribution Vectors}
\label{sec:vectors}

To analyze what fraction of installs is delivered through each distribution
vector we first classify the installer apps as detailed in
Section~\ref{sec:approach}.
Table~\ref{tab:distr_vectors} summarizes the top app distribution vectors we
have identified.  For each distribution vector the left part shows the
percentage of install events (all and unwanted) the distribution vector
is responsible for.  The middle part summarizes the installers in the category:
all installers, 
\malinstallers, 
installers signed with a platform key, and 
package names and signers for the installers. 
\review{
The right part of the table
summarizes the child APKs installed through the distribution vector: number of
packages, signers, and the vector detection ratio.
VDR is the fraction of unwanted APKs installed using a distribution vector 
over all APKs installed via that vector.
RVDR is the relative VDR with respect to \Play, which is set as 1.0.}
Overall, as we explained in Section~\ref{sec:approach}, 
we were able to classify 
14\% of the installers covering 96.3\% of the 412M install events. 
While the fraction of
classified installers is low, we cover the vast majority of installs, enabling
us to accurately compare various distribution vectors.

The main distribution vector is \play, 
responsible for 87\% of all and 67\% of \malinstalls.
While the percentage of \malinstalls is highest for \play, 
its VDR is only 0.6\% and its RVDR the fourth lowest. 
This illustrates that installing from \play is safer than installing from 
most distribution vectors including alternative markets, browsers, and IM.
However, \maldevs have a large incentive to make their apps 
appear in \play since it provides the apps with higher visibility, reputation, and trust. 
This leads to a low fraction, but large number overall, of \malapps 
being able to bypass Play's defenses.
The effectiveness of Play defenses against \malapps is illustrated by 
the lower rate of \malinstalls compared to all installs, 
i.e., they manage to remove a fraction of the \malapps. 
On the other hand, the defenses (if any) against
\malapps used by other distribution vectors, 
save for commercial PPI, do not seem to be effective.
The second largest distribution vector are the over 100 alternative markets
identified, responsible for 5.7\% of all installs and 10.4\% of \malinstalls.
We detail the top 10 markets in Table~\ref{tab:markets} and discuss them below.
Prior work has analyzed the distribution of \malapps through markets 
by crawling official and alternative markets
(e.g.,~\cite{zhou2012hey, wang2018beyond}).
However, such crawling is limited to a fixed set of markets
and a small fraction of apps in each market. 
Also, paid apps are typically ignored.
In addition, some markets 
may not provide a web-based app download interface that researchers can 
easily crawl, e.g., the Vivo market in~\cite{wang2018beyond}.
Compared to prior work, we can observe apps installed by user devices 
regardless of the type of app (paid or free) and from which market they come
(we observe over one hundred alternative markets).

The third distribution vector is through backup restoration.  These installs
correspond to restoration of previously saved apps in the cloud, as well as
transfer of apps while cloning an old phone into a new phone. 
These apps are not an intentional distribution vector, 
but surprisingly they are responsible for nearly 5\% of \malapps installations.
For cloud backups, 
the most likely explanation is that the user decided not to uninstall
the \malapp when prompted by the \client and the app was thus saved. 
This matches with the majority of \malapps installed via this vector being PUP,
for which the \client generates lighter and less frequent notifications.
In some cases it may also happen that the backup was taken before 
installing the \client.
Phone cloning apps are typically privileged (i.e., signed by the platform key) 
so that they can copy all apps in an old phone to the new phone. 
Otherwise, they cannot access the \url{/system/} directory where 
system apps are installed.
Thus, it may happen that privileged \malapps, which cannot be uninstalled 
by the \client, are surviving a phone change by the user thanks to the 
high privilege of the phone cloning apps. 
One example involves a pre-installed \malapp infected with
the \textit{CoolReaper} backdoor~\cite{coolreaper}.
CoolReaper was discovered in phones manufactured by Coolpad, 
a Chinese device vendor, and among its many capabilities, it can
perform fake over-the-air (OTA) updates for installing other \malapps. 
In conclusion, there seems to be an opportunity for backup and phone 
cloning apps to improve defenses against \malapps, 
e.g., by performing AV scans on the saved apps.

\review{Installs by package installers rank fourth by fraction of installs, 
but second by \malinstalls (10.5\%).
These largely correspond to manual installs by the user, 
who may be consciously installing \malapps that offer desired functionality.
For these installs, the vector through which the user downloaded the app 
into the phone is not known.
}

Bloatware is another surprisingly high distribution vector, 
being responsible for 6\% of \malinstalls.
Bloatware are pre-installed apps with unclear functionality.
As explained in Section~\ref{sec:approach} we consider in this category 
apps signed by a device vendor or a carrier, which do not belong to any of the 
other categories, 
i.e., for which we do not understand why they are installing apps.
The most likely reason behind installs in this category, 
as well as the Other and Unclassified categories, is advertising. 
In other words, if we do not understand why an app 
is installing apps from other signers, then we assume that publishers of the 
child apps are paying for the installations.
We discuss such pay-per-install (PPI) agreements in Section~\ref{sec:ppi}.
In summary, this high number of installs by bloatware 
likely indicates aggressive ad-based monetization by 
device vendors and carriers of the phones they sell. %

The browser category shows that app downloads from the Web are rare
(\textless 0.1\% of all installs), 
but have the highest risk of being unwanted (3.8\% VDR). 
In particular, the browser VDR is larger than that of alternative markets. 
Downloading apps through the browser is a riskier 
proposition than downloading them from markets, even the alternative ones. 
This highlights a need for stronger browser-based defenses 
against \malapp downloads.

Next, we analyze the markets and browsers categories to understand 
differences between apps in the same category.

\begin{table}[t]
\caption{Top 10 markets by number of child signers.}
\footnotesize
\centering
\begin{tabular}{|r|l|r|r|r|}
\cline{4-5}
\multicolumn{3}{c|}{} & \multicolumn{2}{c|}{\textbf{Children}} \\
\hline
 \textbf{Rk}
 &\textbf{Market}
 &\textbf{IDR}
 &\textbf{Sig.}
 &\textbf{Pkg.} \\
\hline
	1  & com.android.vending             & 0.6\%   & 816K & 1.2M\\ %
	7  & com.sec.android.app.samsungapps & 1.2\%   & 14K  & 26K \\ %
	8  & com.mobile.indiapp              & 1.6\%   & 12K  & 15K \\ %
	9  & com.amazon.venezia              & 0.7\%   & 12K  & 23K \\ %
	10 & com.oppo.market                 & 2.8\%   & 10K  & 12K \\ %
	11 & com.xiaomi.mipicks              & 1.1\%   & 10K  & 12K \\ %
	12 & com.farsitel.bazaar             & 10.5\%  & 10K  & 20K \\ %
	13 & ir.mservices.market             & 4.4\%   &  8K  & 13K \\ %
	15 & com.vivo.appstore               & 0.9\%   &  8K  & 9K  \\ %
	18 & com.huawei.appmarket            & 11.7\%  &  7K  & 9K  \\ %
\hline
\end{tabular}
\label{tab:markets}
\end{table}

\paragraph{Markets.} Table~\ref{tab:markets} shows the top 10 markets by number
of child signers.  Each row corresponds to an installer, 
i.e., package name and signer pair. The top row corresponds to \Play
(\url{com.android.vending}). It does not include the previous
package name for the official market (\url{com.google.android.feedback}) nor
Play APKs from other signers such as the AOSP test key.
For each market, the table shows the rank by number of child signers 
across all categories. %
We observe significant differences in IDR for different markets.  The highest
IDR of 11.7\% is for the Huawei market (\url{com.huawei.appmarket}), followed
by the Iranian Bazaar market (\url{com.farsitel.bazaar}) with 10.5\%, the
Iranian MyKet market (\url{ir.mservices.market}) with 4.4\%, the NearMe market
from Chinese vendor Oppo (\url{com.oppo.market}) with 2.8\%, and the 9Apps
Indian market (\url{com.mobile.indiapp}) with 1.6\% IDR.
On the better side of the spectrum, there are \play and Amazon's market with
the lowest IDRs of 0.6\% and 0.7\% respectively.  This indicates that the
security vetting process that \play applies to uploaded apps indeed has a
positive effect on user security~\cite{playprotect}.
Compared to \play, the users of alternative markets have up to 19 times
higher probability of encountering \malapps. 

\begin{table}[t]
\caption{Top 10 browsers by number of child signers.}
\footnotesize
\centering
\begin{tabular}{|r|l|l|r|r|}
\cline{5-5}
\multicolumn{4}{c|}{} & \multicolumn{1}{c|}{\bf Child} \\
\cline{1-4}
 \textbf{Rk}
 &\textbf{Browser}
 &\textbf{Name}
 &\textbf{IDR}
 &\textbf{Sig.} \\
\hline
42  & com.UCMobile.intl      & UC             & 3.8\%  & 1,593\\ %
45  & com.android.chrome     & Chrome         & 3.9\%  & 1,521\\ %
80  & com.opera.browser      & Opera          & 3.6\%  & 536\\ %
142 & com.uc.browser.en      & UC Mini        & 5.0\%  & 225\\ %
158 & org.mozilla.firefox    & Firefox        & 3.6\%  & 193\\ %
164 & com.opera.mini.native  & Opera Mini     & 10.5\% & 183\\ %
166 & com.brave.browser      & Brave          & 5.1\%  & 175\\ %
173 & com.coloros.browser    & Oppo ColorOS   & 4.0\%  & 157\\ %
197 & com.android.browser    & Android (Oppo) & 8.8\%  & 133\\ %
215 & com.nearme.browser     & Oppo NearMe    & 6.7\%  & 111\\ %
\hline
\end{tabular}
\label{tab:browsers}
\end{table}

\paragraph{Browsers.} 
Table~\ref{tab:browsers} is similar for the top 10 browsers. 
UC Browser tops the table followed closely by Chrome.
These two browsers rank 42 and 45, respectively, among all installers.
The top seven browsers are available in \play, 
while the last three correspond to 
browsers preinstalled in Oppo phones. 
Similar to the markets, we can observe significant differences in 
IDR between some mobile browsers.  
Most browsers have an IDR in the range 3.8\%--5.1\%, 
but Opera Mini has twice that risk (10.5\% IDR), 
even more compared to the full Opera browser (3.6\%).
We don't have a good explanation for the difference 
between Opera versions, 
as prior work comparing mobile browser security does not flag 
significant differences~\cite{luo2019time}. 

\paragraph{Privileged installers.}
An orthogonal classification is whether installers are system or user level 
apps. 
System-level installers signed by a platform key are responsible for 
4.1\% of all installs and 9\% of \malinstalls. 
The high ratio of \malinstalls is especially worrying 
because these installers have access to system level permissions and 
cannot be uninstalled by normal users or security tools, 
only by the superuser or through ADB.
Security tools can only recommend the user to disable them.
Most system-level installers come pre-installed, 
but we observe that 35\% of their installs are for other system-level apps. 
Thus, it is possible for other system-level installers to be installed 
later in the device lifetime.
Column \textit{Plat.} in Table~\ref{tab:distr_vectors} 
shows the number of system-level installers per category,
which is dominated by bloatware and backup (i.e., phone cloning) apps.
More than half of installers in those two categories are privileged. 
This matches common complaints by users that bloatware is installing 
apps in their phones and cannot be uninstalled.

\paragraph{Considering all installs.}
\review{
So far, we have analyzed the Subset dataset %
corresponding to the 412M install events for which we could 
recover parent information.
Table~\ref{tab:distr_vectors_all} in the Appendix 
presents the same results in Table~\ref{tab:distr_vectors}, 
but for the Full dataset of 1.7B install events.
Results in Table~\ref{tab:distr_vectors_all} could misclassify 
some installs due to impersonation,
but avoid any bias introduced when selecting the Subset dataset.
The results on both datasets are very similar, 
indicating that no significant sampling bias was introduced in the 
Subset dataset.}

\review{
\paragraph{Takeaway.} 
To summarize, we observe that \play is the main app distribution vector 
responsible for 87\% of all installs and 67\% of \malinstalls. 
However, its VDR is only 0.6\%, showing that \play defenses 
against \malapps work, but still significant amounts of \malapps are able 
to bypass them, making it the main distribution vector for \malapps.
Among the remaining installs,
alternative markets are the largest, being  
responsible for 5.7\% of all installs and 10.4\% of \malinstalls. 
Furthermore, on average they are five times riskier (3.2\% VDR) than \Play (0.6\%).
App downloads from the Web are rare (\textless0.1\% installs),
but have a significantly higher risk.
Backup restoration is an unintended \malapp distribution vector 
responsible for 4.8\% of \malinstalls.
Bloatware is another surprisingly high distribution vector, 
responsible for 6\% of unwanted installs. 
}

\subsection{Pay-Per-Install}
\label{sec:ppi}

Pay-per-install (PPI) is a software distribution model 
where an \textit{advertiser} pays \textit{publishers} to 
advertise a program and have it installed on user devices. 
Publishers are paid a commission for 
each confirmed \textit{install} on a new device.
The advertiser can reach direct agreements with publishers, 
e.g., setting up an affiliate network, 
or can leverage \textit{PPI services}, 
who act as middle-men between advertisers and publishers.

Previous work analyzed Windows PPI services, dividing them into 
underground and commercial~\cite{ppi,ppipup,kurt_ppi}.
Underground PPI services mostly distribute malware,
do not advertise themselves publicly, and often use silent installs, 
i.e., the user is unaware of the installation~\cite{ppi}. 
Commercial PPI services, instead, are backed by companies and prompt offers 
for users to decide about the install~\cite{ppipup,kurt_ppi}. 
In both types, publishers are paid for each install of 
a \textit{PPI installer} program
that then downloads the currently advertised programs.

As far as we know, Android PPI services have not yet been analyzed, 
so we provide a first look at them.
Advertising in Android often uses \text{ad libraries} that 
ad networks ask publishers to include into their apps.
For PPI services, if the publishers are owners of popular apps, 
the service may provide them with a library to include in their apps. 
On the other hand, if the publishers are 
device vendors or carriers, the PPI service provides them with a 
stand-alone installer to be pre-installed in their branded devices.
As a starting point, we knew the package names of three stand-alone installers 
for two Android commercial PPI services, 
IronSource (IS) and DigitalTurbine (DT),
mentioned in prior work~\cite{preinstalled}. 
Both commercial PPI services partner with device vendors and 
carriers to pre-install their installer, 
which can then offer apps to the user. 
If the user installs an advertised app, PPI and partner
(carrier or vendor) split the commission paid by the advertiser.
In addition, we had a list of 59 apps identified 
in prior work~\cite{kevincreepware} that pay their users a commission 
if they install other apps. 
In this PPI model, there is no publisher and the PPI service directly 
interacts with users.
While these 59 apps appear in a handful of devices in the reputation logs, 
they did not perform any installs in our dataset.
Starting from the known IS and DT PPI installers, 
we leveraged the 34.6M APKs in the reputation logs 
to identify further installers for those two services. 
For this, we examined apps from the same signer and/or 
similar package name. 
This process identified 48 IS and 38 DT installers. 
The identified 86 installers were used to produce the PPI results in 
Table~\ref{tab:distr_vectors}.

The PPI row in Table~\ref{tab:distr_vectors} shows that those two 
commercial PPI services are responsible for 0.2\% of all installs and 
0.1\% of \malinstalls with DT being responsible for 644K installs 
and IS for 36K. Thus, DT is the larger PPI.
Obviously, this is a very conservative lower bound on 
commercial PPI distribution
as more Android commercial PPI services likely exist. 
However, we have not been able to identify more stand-alone 
commercial PPI installers in our dataset.
It may happen that DT and IS dominate the agreements with carriers and vendors.
Thus, other commercial PPI services provide their publishers with 
a library to embed into their apps, rather than a stand-alone installer. 
Those PPI publishers should then appear under the Other and Unclassified 
categories in Table~\ref{tab:distr_vectors}.
Thus, we can conservatively estimate that all PPI activity 
(commercial and underground) 
can be responsible between 0.1\% and 4\% of all \malapp installs.
Of course, assuming all uncategorized installers are involved in PPI 
is a conservative upper bound as well.
But, that upper bound is already significantly lower 
than the estimate that Windows PPI services distributed 
over a quarter of all PUP~\cite{ppipup}. 
Next, we detail these two commercial PPIs. 
Then, in Section~\ref{sec:malparents} we examine the unknown 
installers that may be involved in PPI distribution.

\paragraph{IronSource.} 
IS is an Israeli advertising company~\cite{ironsource}.
Its offering includes the Aura out-of-the-box experience (OOBE) 
platform, which they claim is installed in 130M devices~\cite{ironsourceAura}. 
The 48 IS installers include the name of vendor and carrier partners 
in the package name
(e.g., \url{com.aura.oobe.samsung}). 
We observe 37 partners: 
29 vendors (e.g., alcatel, huawei, htc, samsung, xiaomi, zte), 
7 carriers (e.g., digicel, hutchinson, telus), and 
one OS publisher (remix).
Of those, 8 vendors (dewav, huawei, irulu, lge, longcheer, tinno, yulong) and
the OS publisher (remix) sign the IS installer with their platform keys,
which gives it system privilege. 
The IDR for IS installers ranges 0\%--5.6\%, 
with a mean of 0.3\% and a median of 0\%.
Thus, IS installs very few \malapps. 
In contrast, their Windows PPI service had a 81\% IDR~\cite{ppipup}. 
The most popular child apps are:
Wish shopping, %
Booking, %
and Candy Crush Soda Saga. %
We also observe TikTok, Netflix, Outlook, SnapChat, Pinterest, Twitter, 
Skype, and Spotify. 
This shows that some of the most popular apps leverage IronSource 
to increase their user base.
To conclude, we observe that IS has significantly cleaned their practices. 
Their vetting limits abusive advertisers, achieving a IDR lower
than other distribution vectors such as markets and browsers.
And, while they use many certificates and package names, 
those are clearly labeled as belonging to IronSource. 

\paragraph{Digital Turbine.}
DT is a public company headquartered in Austin, Texas. 
According to their 2019 fiscal year statement~\cite{dt2019fiscal}
it works with 30 carriers and OEM vendors, is installed in 
260M devices, and has delivered one billion app installs. %
Their revenue for 2019 totaled \$103.6M. 
Similar to IS, the 38 installers include the partner name
(e.g., \url{com.dti.att}). %
We observe 21 partners:
9 carriers (e.g., att, cricket, comcast, uscc), 
18 vendors (e.g., blu, lenovo, samsung, zte), and 
two others (sliide, smartapp).
The IDR for the installers ranges 0.0\%--9.7\% 
with a mean of 1.1\% and a median of 0\%.
This is worse than IS, but still low compared with other sources.
The most downloaded apps are popular:
Facebook, %
Slotomania, and %
Empire: Four Kingdoms. %
We also observe Instagram, Yelp, and YahooMail. 
The advertised apps are mostly disjoint from those IS advertises, 
but we found a few advertised through both PPIs,
e.g., Wish Shopping and Candy Crush Saga.
While the advertised apps are predominantly benign, 
\url{com.dti.gionee} (9.7\% IDR) is a clear exception, 
distributing apps from two advertisers in Table~\ref{tab:malpub} 
(VideoBuddy and MrOwl). 
Gionee mostly sells its devices in India,
thus both apps likely target Indian users.
We also found user reports that the installer name displayed 
in the device was changed from {\it DT Ignite} to {\it Mobile Services Manager},
with users complaining that this was done for obfuscation~\cite{dtNameChange}.
Thus, while their overall IDR is not bad, 
DT can still improve its transparency and advertiser vetting.

\review{
\paragraph{Takeaway.} 
We provide a very conservative lower bound on commercial PPI service
distribution of 0.2\% of all installs and 0.1\% of \malinstalls and observe 
that such services seem to have improved their filtering of abusive 
advertisers compared to their Windows counterparts.
We also estimate that all PPI activity may be responsible for up to 
4\% of the unwanted app installs.
That upper bound is still significantly lower than the estimate of 
Windows commercial PPI services being responsible 
for over a quarter of PUP installs~\cite{ppipup}. 
}

\begin{table*}[t]
\caption{Candidate installers related to PPI identified.
For each installer, it shows 
whether available in Play (GP), 
whether signed by a platform key (Plt.), 
whether some APKs have the system-level INSTALL\_PACKAGES permission (IP), 
SDR, 
IDR, 
number of children packages and signers, and 
the percentage of child packages in Play (GPR).
}
\centering
\resizebox{\textwidth}{!}{
\begin{tabular}{|r|l|l|c|l|r|r|r|r|r|r|}
\cline{9-11}
\multicolumn{8}{c|}{} & \multicolumn{3}{c|}{\bf Children} \\
\hline
 \textbf{\#}
 &\textbf{Package}
 &\textbf{Cert Thumbprint}
 &\textbf{GP}
 &\textbf{Plt.}
 &\textbf{IP}
 &\textbf{SDR}
 &\textbf{IDR}
 &\textbf{Pkg}
 &\textbf{Sig}
 &\textbf{GPR} \\
\hline
1  & cn.feelcool.superfiles             & b0d2737aa9070973f8b66755f9cd32d98fd0bd83 & \N & \N  & \Y & 82.4\% & 71.4\% & 12 & 11 &  0.0\%  \\ %
2  & com.google.android.play.ms72       & 82f0e9ff5dd5ad52cf74eb5e7189a3278ca76358 & \N & \Y  & \N &  0.6\% & 50.0\% & 12 & 12 & 41.7\%  \\ %
3  & com.snaptube.premium               & be135353437d704f3a37e2b413d040a5ddff4f19 & \N & \N  & \N &  0.1\% & 42.6\% & 28 & 30 & 14.3\%  \\ %
4  & cn.opda.a.phonoalbumshoushou       & 8f8360b284a2dfd65dffe47acbd64ffff674cfee & \N & \N  & \Y &  3.3\% & 18.5\% & 23 & 23 & 17.4\%  \\ %
5  & launcher3.android.com.hivelauncher & 28af6c75244a9cbd3f8aee304c425cdc1c66bc6c & \N & \Y  & \Y &  0.0\% & 10.5\% & 30 & 28 & 60.0\%  \\ %
6  & com.vivo.game                      & 283d60ddcd20c56ea1719ce90527f1235ae80efa & \N & \Y  & \Y &  0.4\% & 10.5\% & 18 & 17 &  0.0\%  \\ %
7  & com.zte.aliveupdate                & 1ef46c04828e8994daab682bfe3211cae775a2b4 & \N & \N  & \Y &  1.5\% & 10.4\% & 31 & 23 & 29.0\%  \\ %
8  & com.miui.system                    & 7b6dc7079c34739ce81159719fb5eb61d2a03225 & \N & \Y  & \N &  0.3\% & 9.1\%  & 11 & 11 & 63.6\%  \\ %
9  & com.transsion.appupdate            & 37f3837469049e6022f3248b84372badb77d1a1e & \N & \Y  & \N &  0.5\% & 8.9\%  & 25 & 24 & 84.0\%  \\ %
10 & com.rahul.videoderbeta             & 9816a59361ccd7c33542205da5c7178f32f38042 & \N & \N  & \N &  1.1\% & 8.3\%  & 21 & 21 &  9.5\%  \\ %

\hline
\end{tabular}
}
\label{tab:malparents}
\end{table*}

\subsection{Top Unknown Installers}
\label{sec:malparents}

In this Section we take a look at the installers for which it is not 
clear based on their public description why they would need to install 
other apps, other than for pay-per-install advertising.
These correspond to three rows in Table~\ref{tab:distr_vectors},
the installers that our classification labels as 
Bloatware and Other, as well as those left Unclassified.
To make sure we analyze relevant installers, 
we focus on installers 
that install apps from at least 10 signers.
Table~\ref{tab:malparents} shows the top 10 installers 
satisfying those constraints, sorted by decreasing IDR. 
The top two installers are Unclassified, 
five are Bloatware (ranks 5--9), 
and three are Other,
two video downloaders (Snaptube, Videoder) and 
one optimizer (Baidu Mobile Guard).

The top unknown installer is \url{cn.feelcool.superfiles}. 
It is clearly an \malinstaller since 
all its APKs known to VT are considered unwanted (i.e., VT $\ge$ 4); 
its signing key has a very high SDR (82.4\%),
i.e., its signer mostly signs \malapps; and
it has a very high IDR (71.4\%),
i.e., it mostly installs \malapps. 
Furthermore, none of the 12 apps it installs are available in \Play. 
AVClass labels its samples as adware from the 
\textit{hiddenads} family.
We believe we are the first to cast light on this \malinstaller.
The rest of the installers are not unwanted 
(e.g., low SDR), but install a significant fraction of \malapps.

None of these installers are available in \Play, so we examine how they arrived
in the phone.  We have no installation events for the top two Unclassified
installers, 
Thus, they might come pre-installed, similar to the five bloatware apps.  The
apps in the Other category are installed from multiple alternative markets
where they are quite popular.

Now we investigate whether those installs could have happened without
user consent. 
While we do not know if there was user consent, 
we can determine that in some cases there had to be, 
as the installer lacks the necessary permissions for silent installations.
Column \textit{IP} in the table captures
whether at least one APK from the installer have a VT report stating
that it requests the Android system-level INSTALL\_PACKAGES permission in their
manifest.  This permission is a pre-requisite for performing installs without
user consent.  The results shows that half of the installers have that
permission and thus could perform silent installs if they wanted. 
Note that having the ability to perform silent installs does not mean they 
use it. 
We further discuss this issue in Section~\ref{sec:discussion}. 

\review{
\paragraph{Takeaway.}
The top unknown \malinstallers often distribute apps 
not available in \Play. 
Among these, \url{cn.feelcool.superfiles} clearly stands out 
distributing almost exclusively \malapps not in \Play.
Seven of these installers likely come pre-installed indicating 
PPI agreements with vendors and carriers. 
Of the pre-installed installers, five can perform installations 
without user consent.
However, we do not know whether they use that capability.
}

\section{Related Work}
\label{sec:related}

Few studies have analyzed malware prevalence and distribution on real
Android devices~\cite{preinstalled, kevincreepware, shen2016insights}.
Recently, Gamba et al.~\cite{preinstalled} analyzed pre-installed  apps 
in 2.7K Android devices.
They discover that a significant fraction of pre-installed software
exhibits potentially unwanted behavior like personal data collection
and user tracking.
Our work considers pre-installed bloatware 
as one distribution vector and compares it with other vectors. 
Furthermore, our device dataset is three orders of magnitude larger.
Roundy et al.~\cite{kevincreepware} propose an approach for detecting 
stalking apps.
They evaluate the detection on 50M Android devices, during the period
of 2017--2019, discovering 855 stalking apps in 172K devices.
Our analysis, in comparison, uses a smaller device dataset for a fourth-month period.
However, we analyze a large variety of threats against users.

Several studies have quantified malware in Google Play Store~\cite{viennot2014measurement,carbunar2015longitudinal} 
and third-party markets~\cite{wang2018beyond,ng2014android,petsas2017measurement}.
In comparison, our work measures malware distribution
via multiple channels like \Play store, alternative markets, browsers, IM,
and PPI services.
Prior work has analyzed Windows malware distribution
through underground pay-per-install services~\cite{ppi},
drive-by downloads~\cite{provos2008all,grier2012manufacturing},
free streaming services~\cite{rafique2016s},
and download portals~\cite{rivera2019costly,geniola2017large}.
Results from these studies do not necessarily extrapolate on Android 
due to inherent platform differences.

Prior academic work identifies emerging trends in the Android malware ecosystem~\cite{zhou2012dissecting,suareztangil2018years,felt2011survey,
lindorfer2014andrubis,tang2019large}.
Suarez-Tangil et al.~\cite{tang2019large} conducts 
a behavioral analysis of 1.2M Android malware samples collected from malware feeds,
over a period of eight years (from 2010-2017).
We measure malware families prevalence on real devices 
while prior work uses malware feeds that may be biased towards highly polymorphic families.
Yearly industrial threat reports analyze new threats in
the Android malware ecosystem~\cite{mcafee2019, trendmicro2018, crowdstrike, googlethreat2018}.
A comparison is difficult 
since they measure prevalence by number of detections and not by number of devices.

Other works have analyzed mobile advertising libraries
including information 
leaks~\cite{ad_son2016mobile,demetriou2016free,stevens2012investigating} and
defenses against ad fraud~\cite{ad_shekhar2012adsplit,ad_dong2018frauddroid}.
In our work, we observe a large prevalence of Ad libraries among AV detections. 

\section{Limitations}
\label{sec:discussion}

In this section, we discuss limitations of our work and possible 
avenues for improvement. 

\paragraph{Selection bias.}
Our dataset presents some selection bias worth revisiting. 
First, the reputation logs only include devices with an AV installed. 
Devices without an AV may have a different prevalence of \malapps, 
which we hypothesize may be somewhat larger. 
For example, the \client may block, or alert the user about, 
some \malinstallers,  
reducing the number of \malapps they would otherwise install.
Thus, our prevalence estimates may be conservative when considering all 
mobile devices.
Second, the geographic distribution is skewed towards countries where the 
\vendor has a larger market share. 
We observe some large countries like China, Indonesia, Pakistan, Bangladesh, 
and Nigeria may be underrepresented, 
but we still have tens of thousands of devices in China and 
several thousands in the others.
\review{
Third, to avoid assigning \malinstalls to benign 
installers we focus our distribution vector analysis on the 412M (24\%)
install events 
for which we can recover the parent's signer.
This selection could be biased towards \malinstallers since they have a shorter
TTL that helps in the recovery. 
This could bias VDR absolute numbers. 
To avoid this selection bias, we also present relative VDR values.
In addition, we repeat the distribution vector analysis using the Full 
dataset of 1.7B install events 
(Table~\ref{tab:distr_vectors_all} in the Appendix).
Using the Full dataset some \malinstalls could be wrongly assigned to benign 
installers, but selection bias should be removed. 
Both sets of results are very similar, 
indicating that the Subset dataset has no significant sampling bias.

\paragraph{Pre-installed apps.}
Previous works have considered pre-installed apps to be the ones installed 
under the \url{/system/} directory in an Android device~\cite{preinstalled}. 
Unfortunately, our dataset does not include the APK's installation path.
In this work, we use two proxies to analyze pre-installed apps. 
First, apps signed by the device's platform key are very likely pre-installed. 
Second, we classify as bloatware those apps signed by a carrier or device vendor
for which we cannot identify their goal, 
but observe user reports labeling them as bloatware.
We believe such bloatware is very likely to have been pre-installed.
Still, in both cases we cannot fully guarantee the apps were pre-installed, 
so we are cautious and only say they may be.

\paragraph{User consent.} 
The reputation logs do not directly capture
if an install had user consent.  To understand which installers may
perform installs without user consent, we have examined apps that request
the INSTALL\_PACKAGES permission, a prerequisite for silent installs.
Unfortunately, a request for this permission does not necessarily mean
that the app is performing silent installs. 
One caveat is that the app may have the permission, but may not use it, 
although that could change at any app update. 
Another caveat is that the permission may not be granted, 
e.g., for user-level apps. 
Still, requesting the permission may indicate an interest
in performing silent installs (e.g., in rooted devices).
}

\paragraph{Play presence.}
We only checked the presence of apps in \Play once 
during February 2020.
Some apps may have been available in the past, 
but had since been removed when we queried them, 
making our 24\% estimate a lower bound. 
Prior work has shown that nearly half of all apps are removed %
from Play during a two year period and that privacy violations are 
an important reason for those removals~\cite{wang2018android}. 
Thus, our lower bound may be conservative, especially for \malapps, 
which may be more likely to be removed.
Another caveat is that we only query \Play by package name and do not check 
if the certificate of the app in the market matches the one in our dataset.
Thus, we may say an app was in the market even if there is a certificate 
mismatch.
Our methodology of a single query for each app at a fixed point in time, 
and using solely the package name, is the 
same used to estimate that only 9\% of 
pre-installed apps are present in \Play~\cite{preinstalled}.
Since 87\% of all installs come from Play, 
it makes sense that our prevalence is a larger 24\% as the ratio is 
expected to increase over the device lifetime.

\paragraph{Malware vs PUP.}
Our classification using AVClass shows that 60\% of \malapks are 
PUP and 40\% malware, 
although the prevalence of both classes is almost identical. 
PUP prevalence could in reality be larger 
as AVClass by default considers a sample is malware, 
i.e., a sample is PUP only if enough grayware related keywords 
are found in its labels.
Thus, it could underestimate PUP samples.
Overall, our results highlight the importance of PUP in Android. 
However, as we discuss in Section~\ref{sec:unwanted},
more research is needed into how to cleanly delineate the boundary 
between PUP and malware, especially on Android.

\paragraph{Publisher clustering.}
A publisher could use multiple app signing keys and certificates.
We did not cluster certificates from the same publisher 
because we only had app metadata available.  
Prior approaches have leveraged the Subject DN~\cite{malsign}, 
which in Android can be spoofed. 
Publishers using certificate polymorphism will be identified 
as multiple publishers, possibly resulting in a lower DR.
Developing a technique to cluster signers is an interesting avenue 
for future work.   

\section{Conclusion}
\label{sec:conclusion}
This work performs an analysis on \malapps distribution vectors on Android devices,
including the official and alternative markets, as well as secondary vectors
such as Web downloads, pay-per-install (PPI) services, 
bloatware, backup restoration, and even instant messaging (IM) tools.
We identify that between 10\% and 24\% of users devices encounter at least 
one \malapp.
We reveal that \Play is indeed the main app distribution vector of both benign and \malapps,
while, it has the best defenses against \malapps.
Alternative markets distribute fewer apps but have higher probability to be unwanted.
Bloatware is another surprisingly high distribution vector.
Web downloads are rare and much more risky even compared to alternative markets.
Surprisingly, \malapps may survive users' phone replacement due to
the usage of automated backup tools.
Finally, we observe that app distribution via commercial PPI services
on Android is significantly lower compared to Windows.

\section*{Acknowledgments}
This research was supported by the Regional Government of Madrid 
through grant BLOQUES-CM P2018/TCS-4339 
and by the Spanish Government through the SCUM grant RTI2018-102043-B-I00.
This research received funding from the European
Union’s Horizon 2020 Research and Innovation Programme under Grant Agreement
No. 786669.
Any opinions, findings, and conclusions or recommendations expressed in 
this material are those of the authors or originators, and 
do not necessarily reflect the views of the sponsors.
\platon{Update}
 
{
\bibliographystyle{abbrv}
\bibliography{bibliography/paper}

\begin{thebibliography}{10}

\bibitem{googleids}
{Google Play Advertising ID}.
\newblock
  \url{https://support.google.com/googleplay/android-developer/answer/6048248?hl=en}.

\bibitem{playprotect}
{Google Play Protect}.
\newblock \url{https://www.android.com/play-protect/}.

\bibitem{cc_iso}
Iso 3166-1 alpha-2.
\newblock \url{https://en.wikipedia.org/wiki/ISO_3166-1_alpha-2}.

\bibitem{dtNameChange}
{PSA: disable Mobile Services Manager in System Apps}.
\newblock
  \url{https://forum.xda-developers.com/galaxy-s9-plus/how-to/psa-disable-mobile-services-manager-apps-t3764366}.

\bibitem{coolreaper}
Coolreaper: The coolpad backdoor, 2014.
\newblock
  \url{https://www.paloaltonetworks.com/apps/pan/public/downloadResource?pagePath=/content/pan/en_US/resources/research/cool-reaper}.

\bibitem{judy}
{The Judy Malware: Possibly the largest malware campaign found on Google Play},
  May 2017.
\newblock
  \url{https://blog.checkpoint.com/2017/05/25/judy-malware-possibly-largest-malware-campaign-found-google-play/}.

\bibitem{trendmicro2018}
2018 mobile threat landscape, 2018.
\newblock
  \url{https://www.trendmicro.com/vinfo/in/security/research-and-analysis/threat-reports/roundup/2018-mobile-threat-landscape}.

\bibitem{esetransom2018}
Android ransomware: From android defender to doublelocker, 2018.
\newblock
  \url{https://www.welivesecurity.com/wp-content/uploads/2018/02/Android_Ransomware_From_Android_Defender_to_Doublelocker.pdf}.

\bibitem{crowdstrike}
2019 crowdstrike mobile threat landscape report, 2019.
\newblock
  \url{https://www.crowdstrike.com/blog/mobile-threat-report-2019-trends-and-recommendations/}.

\bibitem{googlethreat2018}
Android security and privacy 2018 year in review, 2019.
\newblock
  \url{https://source.android.com/security/reports/Google_Android_Security_2018_Report_Final.pdf}.

\bibitem{dt2019fiscal}
Digital turbine reports fourth quarter and fiscal full year 2019 results, June
  2019.
\newblock
  \url{https://content.equisolve.net/_35bb1d6ea2a3024c2d067486da749df4/mandalaydigital/news/2019-06-03_Digital_Turbine_Reports_Fourth_Quarter_and_Fiscal__577.pdf}.

\bibitem{mcafee2019}
Mcafee mobile threat report, 2019.
\newblock
  \url{https://www.mcafee.com/enterprise/en-us/assets/reports/rp-mobile-threat-report-2019.pdf}.

\bibitem{androidPriviledgedPermission}
Privileged permission whitelisting, 2020.
\newblock \url{https://source.android.com/devices/tech/config/perms-whitelist}.

\bibitem{aghakhani2020malware}
H.~Aghakhani, F.~Gritti, F.~Mecca, M.~Lindorfer, S.~Ortolani, D.~Balzarotti,
  G.~Vigna, and C.~Kruegel.
\newblock {When Malware is Packin'Heat; Limits of Machine Learning Classifiers
  Based on Static Analysis Features}.
\newblock In {\em Network and Distributed Systems Security Symposium}, 2020.

\bibitem{androzoo}
K.~Allix, T.~F. Bissyand{\'e}, J.~Klein, and Y.~Le~Traon.
\newblock {AndroZoo: Collecting Millions of Android Apps for the Research
  Community}.
\newblock In {\em International Conference on Mining Software Repositories},
  2016.

\bibitem{andow2016study}
B.~Andow, A.~Nadkarni, B.~Bassett, W.~Enck, and T.~Xie.
\newblock {A Study of Grayware on Google Play}.
\newblock In {\em IEEE Security and Privacy Workshops}, 2016.

\bibitem{androidProvisioning}
Android.
\newblock Provision devices.
\newblock
  \url{https://developers.google.com/android/work/play/emm-api/prov-devices}.

\bibitem{androidPackageInstaller}
Android.
\newblock Package installer, 2020.
\newblock
  \url{https://developer.android.com/reference/android/content/pm/PackageInstaller}.

\bibitem{appsGeyser}
Appsgeyser.
\newblock \url{https://appsgeyser.com/}.

\bibitem{aviraPUP}
Avira.
\newblock Potentially unwanted applications.
\newblock \url{https://www.avira.com/en/potentially-unwanted-applications}.

\bibitem{barrera2014baton}
D.~Barrera, D.~McCarney, J.~Clark, and P.~C. Van~Oorschot.
\newblock {Baton: Certificate Agility for Android's Decentralized Signing
  Infrastructure}.
\newblock In {\em ACM Conference on Security and Privacy in Wireless \& Mobile
  Networks}, 2014.

\bibitem{triadaKaspersky}
N.~Buchka and M.~Kuzin.
\newblock Attack on zygote: a new twist in the evolution of mobile threats,
  March 2016.
\newblock
  \url{https://securelist.com/attack-on-zygote-a-new-twist-in-the-evolution-of-mobile-threats/74032/}.

\bibitem{ppi}
J.~Caballero, C.~Grier, C.~Kreibich, and V.~Paxson.
\newblock {Measuring Pay-per-Install: The Commoditization of Malware
  Distribution}.
\newblock In {\em USENIX Security Symposium}, 2011.

\bibitem{carbunar2015longitudinal}
B.~Carbunar and R.~Potharaju.
\newblock A longitudinal study of the google app market.
\newblock In {\em IEEE/ACM International Conference on Advances in Social
  Networks Analysis and Mining}, 2015.

\bibitem{chen2015finding}
K.~Chen, P.~Wang, Y.~Lee, X.~Wang, N.~Zhang, H.~Huang, W.~Zou, and P.~Liu.
\newblock Finding unknown malice in 10 seconds: Mass vetting for new threats at
  the google-play scale.
\newblock In {\em USENIX Security Symposium}, 2015.

\bibitem{crussell2012attack}
J.~Crussell, C.~Gibler, and H.~Chen.
\newblock Attack of the clones: Detecting cloned applications on android
  markets.
\newblock In {\em European Symposium on Research in Computer Security}, 2012.

\bibitem{androidRequestInstall}
E.~Cunningham.
\newblock Making it safer to get apps on android o, August 2017.
\newblock
  \url{https://android-developers.googleblog.com/2017/08/making-it-safer-to-get-apps-on-android-o.html}.

\bibitem{netDragonBaidu}
K.-M. Cutler.
\newblock {Baidu Agrees To Buy Chinese Android App Distributor 91 Wireless For
  \$1.9B}, July 2013.
\newblock
  \url{https://techcrunch.com/2013/07/15/baidu-agrees-to-buy-chinese-android-app-distributor-91-wireless-for-1-9b/}.

\bibitem{demetriou2016free}
S.~Demetriou, W.~Merrill, W.~Yang, A.~Zhang, and C.~A. Gunter.
\newblock Free for all! assessing user data exposure to advertising libraries
  on android.
\newblock In {\em Network and Distributed Systems Security Symposium}, 2016.

\bibitem{ad_dong2018frauddroid}
F.~Dong, H.~Wang, L.~Li, Y.~Guo, T.~F. Bissyand{\'e}, T.~Liu, G.~Xu, and
  J.~Klein.
\newblock {Frauddroid: Automated ad fraud detection for android apps}.
\newblock In {\em ACM Joint Meeting on European Software Engineering Conference
  and Symposium on the Foundations of Software Engineering}, 2018.

\bibitem{felt2011survey}
A.~P. Felt, M.~Finifter, E.~Chin, S.~Hanna, and D.~Wagner.
\newblock A survey of mobile malware in the wild.
\newblock In {\em ACM Workshop on Security and Privacy in Smartphones and
  Mobile Devices}, 2011.

\bibitem{felt2011effectiveness}
A.~P. Felt, K.~Greenwood, and D.~Wagner.
\newblock {The Effectiveness of Application Permissions}.
\newblock In {\em USENIX Conference on Web Application Development}, 2011.

\bibitem{preinstalled}
J.~Gamba, M.~Rashed, A.~Razaghpanah, J.~Tapiador, and N.~Vallina-Rodriguez.
\newblock An analysis of pre-installed android software.
\newblock In {\em IEEE Symposium on Security and Privacy}, 2020.

\bibitem{camScanner}
S.~Gatlan.
\newblock {Trojan Dropper Malware Found in Android App With 100M Downloads},
  August 2019.
\newblock
  \url{https://www.bleepingcomputer.com/news/security/trojan-dropper-malware-found-in-android-app-with-100m-downloads/}.

\bibitem{geniola2017large}
A.~Geniola, M.~Antikainen, and T.~Aura.
\newblock {A Large-Scale Analysis of Download Portals and Freeware Installers}.
\newblock In {\em Nordic Conference on Secure IT Systems}, 2017.

\bibitem{necro}
I.~Golovin and A.~Kivva.
\newblock {An advertising dropper in Google Play}, August 2019.
\newblock \url{https://securelist.com/dropper-in-google-play/92496/}.

\bibitem{googlePUP}
Google.
\newblock Unwanted software policy.
\newblock \url{https://www.google.com/about/unwanted-software-policy.html}.

\bibitem{grier2012manufacturing}
C.~Grier, L.~Ballard, J.~Caballero, N.~Chachra, C.~J. Dietrich, K.~Levchenko,
  P.~Mavrommatis, D.~McCoy, A.~Nappa, A.~Pitsillidis, et~al.
\newblock Manufacturing compromise: the emergence of exploit-as-a-service.
\newblock In {\em ACM Conference on Computer and Communications Security},
  2012.

\bibitem{agentsmith}
A.~Hazum, F.~He, I.~Marom, B.~Melnykov, and A.~Polkovnichenko.
\newblock {Agent Smith: A New Species of Mobile Malware}, July 2019.
\newblock
  \url{https://research.checkpoint.com/2019/agent-smith-a-new-species-of-mobile-malware/}.

\bibitem{androidMarketShare}
A.~Holst.
\newblock {Mobile operating systems' market share worldwide from January 2012
  to December 2019}, January 2020.
\newblock
  \url{https://www.statista.com/statistics/272698/global-market-share-held-by-mobile-operating-systems-since-2009/}.

\bibitem{ironsource}
{IronSource}.
\newblock \url{https://www.ironsrc.com/}.

\bibitem{ironsourceAura}
{IronSource Aura}.
\newblock \url{https://company.ironsrc.com/enterprise-solutions/}.

\bibitem{numAndroidDevices}
T.~Kerns.
\newblock {There are now more than 2.5 billion active Android devices}, May
  2019.
\newblock
  \url{https://www.androidpolice.com/2019/05/07/there-are-now-more-than-2-5-billion-active-android-devices/}.

\bibitem{triadaInjection}
A.~Kivva.
\newblock Everyone sees not what they want to see, June 2016.
\newblock
  \url{https://security.googleblog.com/2019/06/pha-family-highlights-triada.html}.

\bibitem{ppipup}
P.~Kotzias, L.~Bilge, and J.~Caballero.
\newblock {Measuring PUP Prevalence and PUP Distribution through
  Pay-Per-Install Services}.
\newblock In {\em USENIX Security Symposium}, 2016.

\bibitem{enterprise}
P.~Kotzias, L.~Bilge, P.-A. Vervier, and J.~Caballero.
\newblock {Mind your Own Business: A Longitudinal Study of Threats and
  Vulnerabilities in Enterprises}.
\newblock In {\em Network and Distributed Systems Security Symposium}, 2019.

\bibitem{malsign}
P.~Kotzias, S.~Matic, R.~Rivera, and J.~Caballero.
\newblock {Certified PUP: Abuse in Authenticode Code Signing}.
\newblock In {\em ACM Conference on Computer and Communication Security}, 2015.

\bibitem{lindorfer2014andrubis}
M.~Lindorfer, M.~Neugschwandtner, L.~Weichselbaum, Y.~Fratantonio, V.~Van
  Der~Veen, and C.~Platzer.
\newblock Andrubis--1,000,000 apps later: A view on current android malware
  behaviors.
\newblock In {\em International Workshop on Building Analysis Datasets and
  Gathering Experience Returns for Security}, 2014.

\bibitem{luo2019time}
M.~Luo, P.~Laperdrix, N.~Honarmand, and N.~Nikiforakis.
\newblock Time does not heal all wounds: A longitudinal analysis of
  security-mechanism support in mobile browsers.
\newblock In {\em NDSS}, 2019.

\bibitem{malwarebytesPUP}
MalwareBytes.
\newblock How to avoid potentially unwanted programs, February 2016.
\newblock
  \url{https://blog.malwarebytes.com/101/2016/02/how-to-avoid-potentially-unwanted-programs/}.

\bibitem{mcdaniel2012bloatware}
P.~McDaniel.
\newblock Bloatware comes to the smartphone.
\newblock {\em IEEE Security \& Privacy}, 10(4):85--87, 2012.

\bibitem{microsoftPUP}
Microsoft.
\newblock How microsoft identifies malware and potentially unwanted
  applications, March 2020.
\newblock
  \url{https://docs.microsoft.com/en-us/windows/security/threat-protection/intelligence/criteria}.

\bibitem{ng2014android}
Y.~Y. Ng, H.~Zhou, Z.~Ji, H.~Luo, and Y.~Dong.
\newblock Which android app store can be trusted in china?
\newblock In {\em IEEE Annual Computer Software and Applications Conference},
  2014.

\bibitem{nortonGrayware}
Norton.
\newblock What is grayware?, August 2015.
\newblock
  \url{https://uk.norton.com/norton-blog/2015/08/what_is_grayware.html}.

\bibitem{oltrogge2018rise}
M.~Oltrogge, E.~Derr, C.~Stransky, Y.~Acar, S.~Fahl, C.~Rossow, G.~Pellegrino,
  S.~Bugiel, and M.~Backes.
\newblock {The Rise of the Citizen Developer: Assessing the Security Impact of
  Online App Generators}.
\newblock In {\em IEEE Symposium on Security and Privacy}, 2018.

\bibitem{wroba}
A.~Orozco.
\newblock {Trojan looks to ``Wrob'' Android users}, October 2013.
\newblock
  \url{https://blog.malwarebytes.com/cybercrime/2013/10/trojan-looks-to-wrob-android-users/}.

\bibitem{petsas2017measurement}
T.~Petsas, A.~Papadogiannakis, M.~Polychronakis, E.~P. Markatos, and
  T.~Karagiannis.
\newblock Measurement, modeling, and analysis of the mobile app ecosystem.
\newblock {\em ACM Transactions on Modeling and Performance Evaluation of
  Computing Systems (TOMPECS)}, 2(2):1--33, 2017.

\bibitem{provos2008all}
N.~Provos, P.~Mavrommatis, M.~Rajab, and F.~Monrose.
\newblock All your iframes point to us.
\newblock 2008.

\bibitem{rafique2016s}
M.~Z. Rafique, T.~Van~Goethem, W.~Joosen, C.~Huygens, and N.~Nikiforakis.
\newblock It's free for a reason: Exploring the ecosystem of free live
  streaming services.
\newblock In {\em Network and Distributed System Security Symposium}, 2016.

\bibitem{rivera2019costly}
R.~Rivera, P.~Kotzias, A.~Sudhodanan, and J.~Caballero.
\newblock Costly freeware: a systematic analysis of abuse in download portals.
\newblock {\em IET Information Security}, 13(1):27--35, 2019.

\bibitem{kevincreepware}
K.~A. Roundy, P.~B. Mendelberg, N.~Dell, D.~McCoy, D.~Nissani, T.~Ristenpart,
  and A.~Tamersoy.
\newblock The many kinds of creepware used for interpersonal attacks.
\newblock In {\em IEEE Symposium on Security and Privacy}, 2020.

\bibitem{seattleCloud}
Seattlecloud.
\newblock \url{https://seattleclouds.com/}.

\bibitem{avclass}
M.~Sebastian, R.~Rivera, P.~Kotzias, and J.~Caballero.
\newblock Avclass: A tool for massive malware labeling.
\newblock In {\em Research in Attacks, Intrusions, and Defenses}, 2016.

\bibitem{attribution}
S.~Sebastián and J.~Caballero.
\newblock {Towards Attribution in Mobile Markets: Identifying Developer Account
  Polymorphism}.
\newblock In {\em ACM Conference on Computer and Communication Security}, 2020.

\bibitem{ad_shekhar2012adsplit}
S.~Shekhar, M.~Dietz, and D.~S. Wallach.
\newblock Adsplit: Separating smartphone advertising from applications.
\newblock In {\em USENIX Security Symposium}, 2012.

\bibitem{moplusSdk}
S.~Shen.
\newblock {Setting the Record Straight on Moplus SDK and the Wormhole
  Vulnerability}, November 2015.
\newblock
  \url{https://blog.trendmicro.com/trendlabs-security-intelligence/setting-the-record-straight-on-moplus-sdk-and-the-wormhole-vulnerability/}.

\bibitem{shen2016insights}
Y.~Shen, N.~Evans, and A.~Benameur.
\newblock Insights into rooted and non-rooted android mobile devices with
  behavior analytics.
\newblock In {\em Annual ACM Symposium on Applied Computing}, 2016.

\bibitem{ad_son2016mobile}
S.~Son, D.~Kim, and V.~Shmatikov.
\newblock What mobile ads know about mobile users.
\newblock In {\em Network and Distributed Systems Security}, 2016.

\bibitem{stevens2012investigating}
R.~Stevens, C.~Gibler, J.~Crussell, J.~Erickson, and H.~Chen.
\newblock Investigating user privacy in android ad libraries.
\newblock In {\em Workshop on Mobile Security Technologies}, volume~10, 2012.

\bibitem{suareztangil2018years}
G.~Suarez-Tangil and G.~Stringhini.
\newblock Eight years of rider measurement in the android malware ecosystem:
  Evolution and lessons learned, 2018.

\bibitem{tang2019large}
C.~Tang, S.~Chen, L.~Fan, L.~Xu, Y.~Liu, Z.~Tang, and L.~Dou.
\newblock A large-scale empirical study on industrial fake apps.
\newblock In {\em IEEE/ACM International Conference on Software Engineering:
  Software Engineering in Practice (ICSE-SEIP)}, 2019.

\bibitem{taylor2017update}
V.~F. Taylor and I.~Martinovic.
\newblock {To Update or not to Update: Insights from a Two-Year Study of
  Android App Evolution}.
\newblock In {\em ACM Asia Conference on Computer and Communications Security},
  2017.

\bibitem{kurt_ppi}
K.~Thomas, J.~A.~E. Crespo, R.~Rastil, J.-M. Picodi, L.~Ballard, M.~A. Rajab,
  N.~Provos, E.~Bursztein, and D.~Mccoy.
\newblock {Investigating Commercial Pay-Per-Install and the Distribution of
  Unwanted Software}.
\newblock In {\em USENIX Security Symposium}, 2016.

\bibitem{viennot2014measurement}
N.~Viennot, E.~Garcia, and J.~Nieh.
\newblock A measurement study of google play.
\newblock In {\em ACM International Conference on Measurement and Modeling of
  Computer Systems}, 2014.

\bibitem{vt}
{VirusTotal}.
\newblock \url{http://www.virustotal.com/}.

\bibitem{wang2018android}
H.~Wang, H.~Li, L.~Li, Y.~Guo, and G.~Xu.
\newblock {Why are Android Apps Removed from Google Play? A Large-Scale
  Empirical Study}.
\newblock In {\em IEEE/ACM International Conference on Mining Software
  Repositories}, 2018.

\bibitem{wang2018beyond}
H.~Wang, Z.~Liu, J.~Liang, N.~Vallina-Rodriguez, Y.~Guo, L.~Li, J.~Tapiador,
  J.~Cao, and G.~Xu.
\newblock Beyond google play: A large-scale comparative study of chinese
  android app markets.
\newblock In {\em Internet Measurement Conference}, 2018.

\bibitem{wp2Apk}
Wordpress2apk.
\newblock \url{https://wp2apk.com/index.php}.

\bibitem{yen2014epidemiological}
T.-F. Yen, V.~Heorhiadi, A.~Oprea, M.~K. Reiter, and A.~Juels.
\newblock {An Epidemiological Study of Malware Encounters in a Large
  Enterprise}.
\newblock In {\em ACM Conference on Computer and Communications Security},
  2014.

\bibitem{zheng2014droidray}
M.~Zheng, M.~Sun, and J.~C. Lui.
\newblock Droidray: a security evaluation system for customized android
  firmwares.
\newblock In {\em ACM Symposium on Information, Computer and Communications
  Security}, 2014.

\bibitem{zhou2012detecting}
W.~Zhou, Y.~Zhou, X.~Jiang, and P.~Ning.
\newblock Detecting repackaged smartphone applications in third-party android
  marketplaces.
\newblock In {\em ACM Conference on Data and Application Security and Privacy},
  2012.

\bibitem{zhou2012dissecting}
Y.~Zhou and X.~Jiang.
\newblock Dissecting android malware: Characterization and evolution.
\newblock In {\em IEEE Symposium on Security and Privacy}, 2012.

\bibitem{zhou2012hey}
Y.~Zhou, Z.~Wang, W.~Zhou, and X.~Jiang.
\newblock Hey, you, get off of my market: detecting malicious apps in official
  and alternative android markets.
\newblock In {\em Network and Distributed Systems Security Symposium}, 2012.

\bibitem{zhu2020measuring}
S.~Zhu, J.~Shi, L.~Yang, B.~Qin, Z.~Zhang, L.~Song, and G.~Wang.
\newblock {Measuring and Modeling the Label Dynamics of Online Anti-Malware
  Engines}.
\newblock In {\em USENIX Security Symposium}, 2020.

\end{thebibliography}

}

\appendix

\begin{figure}
	\includegraphics[width=.95\columnwidth]{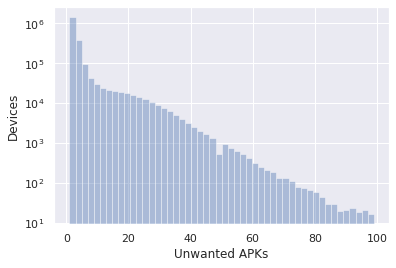}
	\caption{\Malapks per device, for devices with at least one encounter. 
  The distribution is cut at x=100.}
  \label{fig:malapps_per_device}
\end{figure}

Figure~\ref{fig:malapps_per_device} shows the 
distribution of \malapks per device, for devices with at least one encounter.
The median is 2.0, the average 5.0, and the 
standard deviation 1497.0. 
The y-axis is logarithmic and the x-axis s cut at x=100.

\section{Additional Results}
\label{sec:additionalResults}

\review{
\begin{table*}[t]
\small
\caption{\review{Summary of app distribution including installation events without parent signer information.}}
\centering
\begin{tabular}{|l|r|r|r|r|r|r|r|}
\cline{2-6}
\multicolumn{1}{c|}{} & \multicolumn{2}{c|}{{\bf Installs}} & \multicolumn{1}{c|}{\bf Installer} & \multicolumn{2}{c|}{\bf Children} & \multicolumn{2}{c}{} \\
\hline
 \textbf{Vector}
 &\textbf{All}
 &\textbf{Unw.}
 &\textbf{Pkg.}
 &\textbf{Pkg.}
 &\textbf{Sig.}
 &\textbf{VDR}
 &\textbf{RVDR}\\
\hline
\hline
Playstore      & 92.1\%           & 75.9\%          & 2     & 2.3M & 1.5M  & 0.6\% & 1.0 \\
Alt-market     & 2.7\%            & 8.0\%           & 88    & 208K & 125K  & 3.9\% & 6.5 \\
Backup         & 0.6\%            & 1.2\%           & 31    & 595K & 397K  & 0.9\% & 1.5 \\
Pkginstaller   & 1.0\%            & 11.6\%          & 11    & 527K & 332K  & 2.1\% & 3.5 \\
Bloatware      & 0.3\%            & 1.8\%           & 38    &   3K & 1.7K  & 1.3\% & 2.1 \\
PPI            & 0.3\%            & 0.2\%           & 20    &   2K & 1.7K  & 0.3\% & 0.5 \\
Fileshare      & \textless0.1\%   & \textless0.1\%  & 13    & 10K  & 8.7K  & 1.2\% & 2.0 \\
Themes         & \textless0.1\%   & \textless0.1\%  & 2     & 1.3K &   16  & 0.1\% & 0.1 \\
Browser        & \textless0.1\%   & 0.1\%           & 40    & 9.8K & 6.9K  & 3.6\% & 6.0 \\
MDM            & \textless0.1\%   & \textless0.1\%  & 7     & 789  & 506   & 0.3\% & 0.5 \\
Filemanager    & \textless0.1\%   & \textless0.1\%  & 32    & 19K  & 14K   & 2.3\% & 3.8 \\
IM             & \textless0.1\%   & \textless0.1\%  & 10    & 2K   & 1.2K  & 2.8\% & 4.6 \\
\hline
Other          & \textless0.1\%   & 0.2\%           & 127   & 16K   & 10.7K   & 3.5\% & 5.8 \\
Unclassified   & 3.0\%            & 0.9\%           & 79K   & 283K  & 109.5K  & 0.8\% & 1.3 \\
\hline
All            & 100.0\%          & 100.0\%         & 79K   & 2.8M  & 1.8M    & 1.7\% & 2.8 \\
\hline
\end{tabular}
\label{tab:distr_vectors_all}
\end{table*}

Table~\ref{tab:distr_vectors_all} shows the summary of app distribution 
for all 1.7B installation events for which we could identify 
an installation vector, independent of the availability of parent information
(i.e., installer hash and installer signer).
The changes in the VDR column, compared to Table~\ref{tab:distr_vectors},
shows minimal changes.
The largest change is in the VDR value of the alternative markets that increases
by 0.7\% and the unclassified apps vector by 0.7\%. %
The rest are increased at most by 0.3\%.
}

\end{document}